\documentclass[sigconf]{acmart}
\usepackage{graphicx, caption, subcaption}
\usepackage{amsmath}
\usepackage{color}
\usepackage{multirow}
\usepackage{booktabs}
\usepackage{bbm}
\usepackage{enumitem}
\usepackage[flushleft]{threeparttable}
\usepackage{makecell}
\usepackage{hyperref}

\AtBeginDocument{%
  \providecommand\BibTeX{{%
    \normalfont B\kern-0.5em{\scshape i\kern-0.25em b}\kern-0.8em\TeX}}}

\copyrightyear{2022}
\acmYear{2022}
\setcopyright{acmcopyright}\acmConference[SIGIR '22]{Proceedings of the 45th International ACM SIGIR Conference on Research and Development in Information Retrieval}{July 11--15, 2022}{Madrid, Spain}
\acmBooktitle{Proceedings of the 45th International ACM SIGIR Conference on Research and Development in Information Retrieval (SIGIR '22), July 11--15, 2022, Madrid, Spain}
\acmPrice{15.00}
\acmDOI{10.1145/3477495.3532040}
\acmISBN{978-1-4503-8732-3/22/07}

\newcommand{\secref}[1]{Section \ref{#1}}
\newcommand{\figref}[1]{Figure \ref{#1}}
\newcommand{\eqnref}[1]{Eq. (\ref{#1})}
\newcommand{\tabref}[1]{Table \ref{#1}}

\newcommand{\cut}[1]{}

\newcommand{\tabincell}[2]{\begin{tabular}{@{}#1@{}}#2\end{tabular}}

\begin{document}
\fancyhead{}

\title{Positive, Negative and Neutral: Modeling Implicit Feedback in 
Session-based News Recommendation}

\author{Shansan Gong}
\affiliation{%
 \institution{Shanghai Jiao Tong University}
 \streetaddress{800 Dongchuan Rd}
 \city{Shanghai}
 \postcode{200240}
 \country{China}}
\email{gongshansan@sjtu.edu.cn}

\author{Kenny Q. Zhu}
\authornote{Kenny Q. Zhu is the corresponding author and is 
partially supported by NSFC Grant No. 91646205, 
SJTU-CMBCC Joint Research Scheme, and SJTU-Meituan Joint
Research Scheme.}
\affiliation{%
  \institution{Shanghai Jiao Tong University}
  \streetaddress{800 Dongchuan Rd}
  \city{Shanghai}
  \postcode{200240}
  \country{China}}
\email{kzhu@cs.sjtu.edu.cn}


\begin{abstract}
  News recommendation for anonymous readers is a useful but challenging task for many news portals, where interactions between readers and articles 
  are limited within a temporary login session. 
  Previous works tend to formulate session-based recommendation as 
  a next item prediction task, while they neglect the implicit feedback from 
  user behaviors, which indicates what users really like or dislike. 
  Hence, we propose a comprehensive framework to model user behaviors
  through positive feedback (i.e., the articles they spend more time on) and
  negative feedback (i.e., the articles they choose to skip without clicking in).
  Moreover, the framework implicitly models the user using their session start time, and the article using its initial publishing time, in what we call ``neutral
  feedback''.
  Empirical evaluation on three real-world news datasets shows 
  the framework's promising performance of more accurate, diverse and even 
  unexpectedness recommendations than other state-of-the-art 
  session-based recommendation approaches.
\end{abstract}



\maketitle
\section{Introduction}
Online news portals such as BBC, CNN, and Bing News have a huge number 
of readers daily. Many of them are anonymous or logged in as guests who 
typically do not read many stories in a single login session.
Given the limited interactions users engage with the portals, it is often 
hard for the systems to fully understand the user behaviors, 
posing significant challenges to recommendation systems. 

Conventional news recommendation approaches tend to formulate the 
recommendation task as CTR prediction task, and they mainly rely on 
collaborative filtering and factorization machine~\cite{cheng2016wide,guodeepfm2017,wang2018modeling,ge2020graph,hu2020graph,xie2020deep}, 
which requires the system to keep track of the user history 
and can not be applied to anonymous visits or guest logins. 
Recent neural approaches for news recommendation mostly focus on 
encoding the text feature of articles with 
attention mechanism~\cite{wang2018dkn,zhu2019dan,wu_neural_2019-1,wu2019npa,wang2020fine,wu2020CPRS} 
when modeling the user interest while paying little attention to the click 
behavior or the article-to-article transition. 
For example, they have not taken full advantage of the temporal information 
associated with the reading behavior, which is important
especially when the interactions with the user are sparse.

\begin{figure}[th]
    \centering
    \includegraphics[width=0.9\columnwidth]{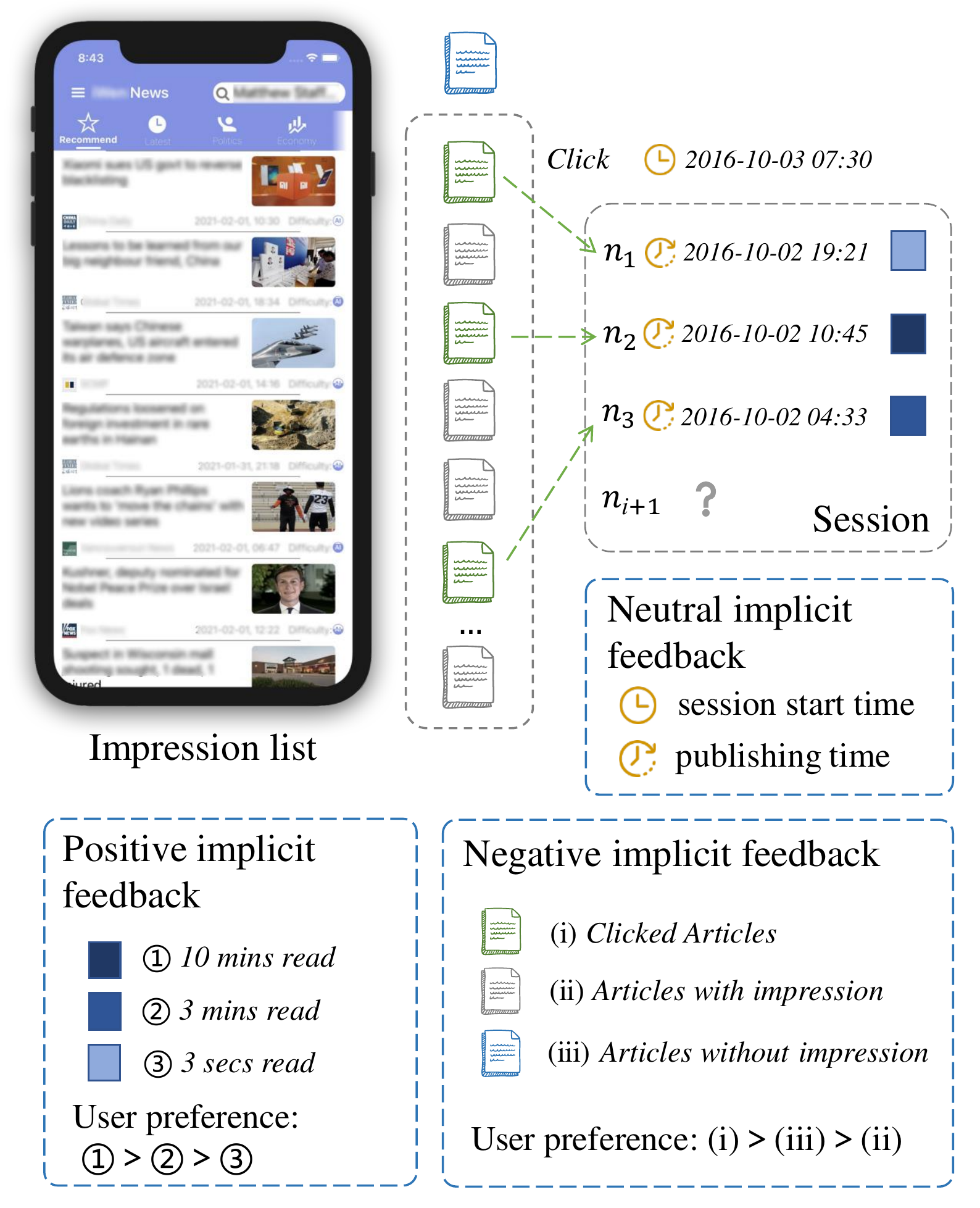}
    \caption{In session-based news reading, 
    a user may spend different amounts of time on different clicked articles, 
representing different level of preference in an implicit form of 
\textbf{positive feedback}; 
a user's impression of an article without an eventual click on the
article represents an implicit form of \textbf{negative feedback}; 
the start time of the click and publishing time of articles can be viewed 
as \textbf{neutral feedback}.}
\label{fig:session}
\end{figure}

Considering the above issues, it's natural and realistic to formulate the 
streaming-like news recommendation task for anonymous users as 
a session-based recommendation task~\cite{sottocornola2018session,gabriel2019contextual,zhang2019dynamic,zhang2018deep,symeonidis2021session}. 
The task is to recommend the next item that the user might be interested 
in given the previous sequence of behaviors within a session, 
where a \textit{session} is usually 
a short period of time (e.g., 30 minutes) during which the user is logged on.
Session-based recommendation is widely used in the e-commerce or 
video streaming domain~\cite{xu2019time,pan2020star}, and can successfully capture
users' short-term intention transition process~\cite{epure_recommending_2017,symeonidis2020session}. However, they rarely consider the implicit feedback from user behaviors.

In this paper, we are interested in exploiting user actions outside the 
clicks themselves. We call them ``implicit feedback''\footnote{To avoid misconception, here the term ``feedback'' is interchangeable with ``signals''.}, as
illustrated in \figref{fig:session}. 
Typical implicit feedback can be extracted from browsing the main page, 
reading an article, closing an article, backtracking~\cite{smadja_understanding_2019}, 
etc. We believe that modeling such implicit feedback ``explicitly'' 
in the session-based recommendation system 
can help the recommender understand user intention better. 
In this work, we focus on answering these questions:
\begin{itemize}
    \item If a user clicked an article, did she really \textit{like} it? 
    \item If a user did not click an article, did she \textit{dislike} it?
    \item How do we model the temporal characteristics of the user and the articles in 
the system?
\end{itemize}

First, in traditional recommendation systems, ``clicks'' usually indicate a ``like'' or a
vote from the user, but things are a bit different for news reading. 
Users may be ``tricked'' into clicking an article~\cite{wang2020click} 
and once they realize that, they will quickly back out and switch to other articles. 
Thus the time a user spends on reading an article is a better, finer-grained 
gauge of the user's preference for the article~\cite{wu2020CPRS}, than 
just the click alone, which is only binary. We model this as the \textbf{implicit 
positive feedback} in this paper. 

Second, just because the user did not click on an article does not necessarily
mean the user does not like it; maybe she was never exposed to this article!  
We can infer what articles might have an impression~\cite{xie2020deep} on 
the user during a session by by assuming that articles are presented to the user 
roughly in the order of their publication time. 
Only those articles within her list of impressions but not clicked are considered
``not interesting'' to her. This is called \textbf{implicit negative feedback}.

Finally, while the positive and negative feedback helps us estimate the 
connection between the user and articles, 
some critical temporal information is useful to model
the user and the articles individually. 
The session start time of a user may suggest
the daily routine of that user. We can expect users who read on
the same day of a week or same time of a day to have to
share the same reading behavior or even background.
On the other hand, the publishing time of each article can also be
formed into a sequence in a session, which reflects the user's sensitivity of 
the timeliness of the stories. We thus carefully design the representation of
session start time and article publishing time as 
\textbf{implicit neutral feedback}.

In this paper, we formulate a session-based recommendation task to 
predict the next possible article 
in each session from the candidate articles pool. Our main contributions are:
\begin{itemize} 
\item For the first time, we leverage the positive/negative/neutral implicit feedback 
in anonymous session-based news recommendation (\secref{sec:approach});
\item We design novel representations for temporal information 
and incorporate it with positive and negative feedback 
into a deep attention network;
\item Our comprehensive offline evaluations on three real-world datasets 
show the clear advantage of our proposed method in terms of overall performance on 
diversity, accuracy and serendipity in both normal and 
cold-start scenarios (\secref{sec:experiment}).
\end{itemize}

\section{Task Definition}
\label{sec:task}
Assume that an anonymous user $u$ produces a sequence of click
events $S_u$ with length $T$, which can be represented as \[S_u: (n_1^u)\rightarrow(n_2^u)\rightarrow...\rightarrow(n_i^u)\rightarrow...,\]
where $n_i^u$ denotes the id of the $i$-th clicked article. 

In the training phase, the recommendation system aims to model the 
user feedback sequence as vector $\mathbf{x_s}$, 
maximizing the similarity between
the vector of predicted next-article ($n_{i+1}^u$) and $\mathbf{x_s}$. 
While in the test phase, 
given interaction sequence $S_u$ as input, 
we produce a ranking list $R$ with the probability that the target user $u$ is likely to click next from high to low. 
Typically a recommender needs to recommend more
than one item for a user, thus the top-k items in $R$ are recommended.
Note that in our anonymous settings, each user $u$ only appears once, 
which means training set and test set do not share the same users. 

\tabref{tb:note} summarizes some critical symbols and notations we use, 
and in later sections, for brevity, we ignore the superscript $u$ 
when discussing the information for a particular user. To explain further, the set $Imp_u$ denotes the articles in the vicinity of the articles
being clicked, which may have left an impression on the user.

\begin{table}[t]
    \caption{Notations and descriptions}
    \begin{tabular}{l|l}
    \toprule
    Symbol & Description\\
    \midrule
    $S_u$ & A session produced by an anonymous user $u$.\\
    $T$ & The number of articles in $S_u$.\\
    $i$ & $i\in [1,T]$ indexes articles in $S_u$.\\ 
    $n_i$ & The id of $i$-th article clicked by $u$.\\
    $\mathbf{c}_i$ & The Content representation of the article $n_i$ in $S_u$ \\
    $t_i$ & The active time that $u$ stays in the article $n_i$. \\
    $\mathbf{ta}_i$ & The encoded active duration that $u$ stays in $n_i$. \\
    $\mathbf{ts}_i$ & The encoded click time that $u$ click for article $n_i$. \\
    $\mathbf{tp}_i$ & The encoded publishing time of article $n_i$. \\
    $Imp_u$ & A set of articles that appear in $u$'s impression list.\\
    $\mathbf{x}_i$ & The item embedding vector of $n_i$.\\
    $\mathbf{xc}_i$ & The embedding vector of $n_i$ combined  $\mathbf{x}_i$ and $\mathbf{c}_i$.\\
    $\mathbf{xc_s}$ & The contextual session vector of $S_u$.\\
    $\mathbf{xt_s}$ & The temporal session vector of $S_u$.\\
    $\mathbf{x_s}$ & The final session vector of $S_u$.\\
    $N$ & The total number of articles.\\
    $j$ & $j\in [1,N]$ indexes total $N$ articles and is article id.\\
    $\hat{y_j}^u$ & The score of article $j$ clicked by $u$ in the next step.\\
    $R$ & Ranking list generated according to $\hat{y_j}^u$.\\
    $Ne_u$ & The set of negative samples for each $S_u$.\\
    \bottomrule
    \end{tabular}
    \label{tb:note}
\end{table}



\section{Approach}
\label{sec:approach}
\begin{figure*}[th]
    \centering
    \includegraphics[width=0.95\textwidth]{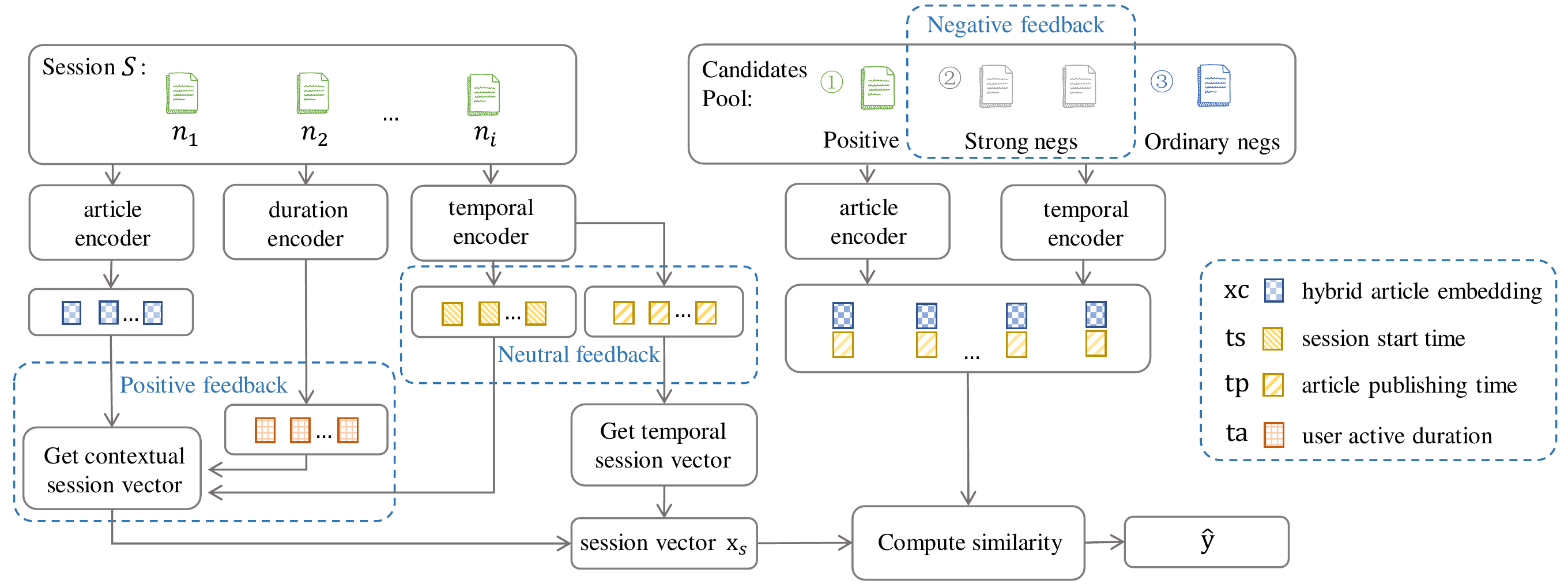}
    \caption{The architecture of our model. Squares in the figure represent
vectors, and their colors refer to the different encoders that produce them.}
    \label{fig:arch}
\end{figure*}

We first lay down some foundations for session-based recommendation, 
then present our base model which is a content-aware session-based model.
After that, we introduce the key ideas of neutral, positive, 
and negative feedback,
which are additional mechanisms that strengthen the base model. 
The high-level overview of our framework is illustrated in \figref{fig:arch}. 

\subsection{Session-based Recommendation Basics}
In a typical session-based setting, given the prefix sequence of the session, denoted as $S_u=(n_1, n_2,...,n_T)$, our goal is 
to predict $n_{T+1}$ article that the target user $u$ is most likely to click next. 
Following~\cite{liu2018stamp}, they use an $N\times d_n$ 
item embedding matrix, where $d_n$ is the embedding dimensionality, to provide article $n_i$'s embedding vector
as $\mathbf{x}_i$. Then methods like RNN~\cite{hidasi2018recurrent}, GNN~\cite{wu2019session,pan2020star}, 
or attention-based approaches~\cite{kang_self-attentive_2018,liu2018stamp} can be used to 
encode the session information into vector $\mathbf{x_s}$ from the sequence 
${(\mathbf{x}_1, \mathbf{x}_2, ..., \mathbf{x}_T)}$, which represents the user's history preferences.
Meanwhile, the same item embedding matrix can be also regarded as $N$ encoded candidates $[\mathbf{x}_1, \mathbf{x}_2,.., \mathbf{x}_N]$.
For $u$, the cosine similarity score $\hat{z_j}^u$ between the session representation and the article $j$ is calculated by the inner product of the session vector $\mathbf{x_s}$ and the candidate news article's vector:
\begin{equation}
    \label{eq:zj}
    \hat{z_j}^u = \mathbf{x}_j^T\mathbf{x_s}, j\in[1,N],
\end{equation}
\begin{equation}
    \label{eq:yy}
    \hat{\mathbf{y}}^u = softmax(\hat{\mathbf{z}}^u),
\end{equation}
$\hat{y_j}^u$ is normalized by softmax function in \eqnref{eq:yy} to be the probability of the article $j$ being clicked next in the session. 
The cross-entropy is usually used to compute loss:
\begin{equation}
    \label{eq:l1}
    \mathcal{L}_1 = - \frac{1}{|S|}\sum_{S_u \in S}\sum_{j=1}^N ( y_j^u \log(\hat{y_j}^u) + (1-y_j^u)\log(1-\hat{y_j}^u)),
\end{equation}
where $S$ is the whole training sessions, $y_j^u=1$ if the article $j$ is indeed the next-clicked articles $n_{T+1}$ in $S_u$ and $y_j^u=0$ otherwise.

\subsection{Our Base Model: Content-aware Recommendation (CAR)}
\label{sec: base}
In order to recommend new and emerging articles,
our starting point is a basic content-aware recommendation model. 
To encode articles' content information, 
Some use pre-trained article content embeddings~\cite{gabriel2019contextual}, 
which is supervised trained based on the Word2Vec word embeddings in news titles and the metadata attributes of articles, such as categories.
Specifically, we get the $d_c$-dimensional vectors from Word2Vec to represent the 
topic-oriented content of articles. Once we get the content vector $\mathbf{c}_i$ of 
article $n_i$, we concatenate $\mathbf{c}_i$ and $\mathbf{x}_i$ to represent 
the article as $\mathbf{xc}_i$. 
To model the varying user preference to the articles in the same session, mainly following~\cite{liu2018stamp}, we adopt a simple attention network, using the weighted sum of the input sequence of vectors to encode the whole session.

We define $\alpha_i$, the attention weight of $i$-th articles $n_i$ in session $S_u$ as:
\begin{equation}
    \label{eq:alpha}
    \alpha_i = W_0 \times \sigma (W_1 \times \mathbf{xc}_i +  b_0),
\end{equation}
\begin{equation}
    \alpha_i^{\prime} = \frac{exp(\alpha_i)}{\sum_{i=1}^T exp(\alpha_i)},
\end{equation}
where $W_0\in \mathbb{R}^{1 \times d_n}, W_1 \in \mathbb{R}^{d_n\times (d_n+d_c)}$ 
are weighting parameters, $\mathbf{xc}_i$ is the vector representing the article, 
and $b_0\in \mathbb{R}^{d_n}$ is a bias. $\alpha_i^{\prime}$ is normalized by softmax function.

Finally, the contextual session vector $\mathbf{xc_s}$ of session $S_u$ is defined as the weighted sum:
\begin{equation}
    \label{eq:final_repre}
    \mathbf{xc_s} = \sum_{i=1}^{T} \alpha_i^{\prime} \mathbf{xc}_i.
\end{equation}

Noted that in order to obtain the sequential information of the input sequence, the attention mechanism extract $\mathbf{xc_s}$ utilizing the positional information by means of the positional encoding, which is the same as it is in Transformer architecture~\cite{vaswani2017attention}. In the end, we can optimize the loss function according to \eqnref{eq:zj}\textasciitilde\eqnref{eq:l1}; when computing $\mathcal{L}_1$, the $\mathbf{x}_j$ in \eqnref{eq:zj} should be replaced by $\mathbf{xc}_j$ and $\mathbf{x}_s$ is equal to $\mathbf{xc_s}$ in this base model.

\subsection{Modeling Time as Neutral Feedback}
\label{sec: temporal}
Active time that one user stays on one particular article is the duration time. For the different and continuous active time $t_i$, we design a duration encoder, which
bucketizes $t_i$ into a discrete variable~\cite{wu2020CPRS} by: 
\begin{equation}
   t_i^{\prime}=\lfloor log_2{t_i} \rfloor,
\end{equation}
and we map discrete $t_i^{\prime}$ into $m$ distinct categories, 
where each category shares the same embedding vector $\mathbf{ta}_i$.

Different from the duration time which reveals the positive implicit feedback 
from users, the date-time when a certain event happens carries physical 
meaning and also conveys the regular pattern behind the temporal information, 
hence we design a temporal encoder for this kind of temporal representation. We encode the publishing time of an article and the start time of a session. In order to extract periodic temporal change which is also known as seasonality, we feed 
the month, day of the month, day of the week, hour, minute $(s\in \mathbb{R}^{12}, d \in \mathbb{R}^{31}, w \in \mathbb{R}^7, h\in \mathbb{R}^{24}, m\in \mathbb{R}^{60})$ into a time embedding matrix $\mathbf{E_t}\in \mathbb{R}^{134 \times d_t}$ with $d_t$ dimension, and concatenate five parts as $\mathbf{t} \in \mathbb{R}^{5d_t}$ to build the whole temporal vector and to represent the date time. We next introduce how to utilize start time and publishing time, then we focus on modeling positive and negative implicit feedback which both derive from temporal information.

\subsubsection{Start time}
Users who start reading at a similar time are more likely to share the 
same reading behavior, which means that user interests are influenced by 
the start time. For example, some people tend to read financial news in 
the morning but instead read entertainment in the evening. 
We denote the click time from each click behavior of a session 
as $\mathbf{ts_i}\in \mathbb{R}^{2d_t}$ using the week and the hour $(w, h)$, which is enough to capture different user's daily routine. To model the different informativeness of the articles in $S_u$ for users' reading at different start reading time, we apply this information to compute personalized attention. We first transform the start time embedding vector $\mathbf{ts_i}$ into a preference query $\mathbf{q_i}$, which is similar to the ``query'' part in Transformer architecture:
\begin{equation}
    \mathbf{q_i} = ReLU(W_t \times \mathbf{ts}_i + b_t),
\end{equation}
where $W_t \in \mathbb{R}^{d_n \times 2d_t}$ is the projection parameter, $b_t \in \mathbb{R}^{d_n}$ is a bias.

Then we evaluate the importance of the interactions between preference query $\mathbf{q_i}$ and article representation $\mathbf{c}_i$ as attention $\alpha^{t}$:
\begin{equation}
    \alpha_i^{t} = \mathbf{c}_i \times tanh ( W_t^{\prime} \times \mathbf{q_i} + b_t^{\prime}),
\end{equation}
\begin{equation}
    \alpha_i^{t\prime} = \frac{exp(\alpha_i^{t})}{\sum_{j=1}^{|S_u|}exp(\alpha_j^{t})},
\end{equation}
where $W_t^{\prime} \in \mathbb{R}^{(d_c+d_n)\times d_n}$ and $b_t^{\prime} \in \mathbb{R}^{d_c+d_n}$ are weighting parameters. The contextual vector representation in \eqnref{eq:final_repre} is now modified to:
\begin{equation}
    \mathbf{xc_s} = \sum_{i=1}^{|S_u|} (\alpha_i^{\prime}+\alpha_i^{t\prime}) \mathbf{xc}_i.
\end{equation}

\subsubsection{Publish time}
Users' reading habits are reflected in the sequence of publishing time ${tp_1,...,tp_i}$ in $S_u$. We can make inferences whether the user tends to browse new articles or older ones from this. The publishing time of clicked articles is a relatively independent sequence thus we model it separately. Due to the high density of article publishing time, we construct publishing time embedding vector $\mathbf{tp}_i\in \mathbb{R}^{5d_t}$ using $(s, d, w, h, m)$. We obtain the session temporal representation vector $\mathbf{xt_s}$ by applying a similar attention mechanism in \secref{sec: base}. We add the content vector of each article to capture the attention relation between the article content and its publishing time. The attention weight with click-level content information involved is formulated as:
\begin{equation}
    \alpha_i^{tp} = W_0^{\prime} \times \sigma (W_1^{\prime} \times \mathbf{tp}_i + W_2^{\prime} \times \mathbf{c}_i + b_0^{\prime}),
\end{equation}
\begin{equation}
    \alpha_i^{tp\prime} = \frac{exp(\alpha_i^{tp})}{\sum_{j=1}^{|S_u|}exp(\alpha_j^{tp})},
\end{equation}
where $W_0^{\prime}\in \mathbb{R}^{1 \times d_n}, W_1^{\prime} \in \mathbb{R}^{(5d_t)\times d_n}, W_2^{\prime} \in \mathbb{R}^{d_c\times d_n}$ and $b_0^{\prime} \in \mathbb{R}^{1 \times d_n}$. The final temporal session representation is:
\begin{equation}
    \mathbf{xt_s} = \sum_{i=1}^{|S_u|} \alpha_i^{tp\prime} \mathbf{tp}_i.
\end{equation}

In the end, we concatenate $\mathbf{xt_s}$ and $\mathbf{xc_s}$ as the aggregated representation $\mathbf{x_s} \in \mathbb{R}^{d_n+d_c+5d_t} $ for the whole session. As for computing $\mathcal{L}_1$, the $\mathbf{x}_j$ in \eqnref{eq:zj} should be replaced by $\mathbf{xc}_j \oplus \mathbf{tp}_j$ ($\oplus$ stands for the concatenation operation).

\subsection{Modeling Positive Feedback}
\label{sec:positive feedback}
Our implicit positive feedback takes the form of the \textit{active time} interval that 
a user spent on each article after clicking on it. If the user spends a short time 
in an article, it's probably because the user is fooled by the title but actually does not like 
the article~\cite{lu_quality_2019}. Note that if the active time is not explicitly available, 
it can be estimated by the time interval between the user's two consecutive clicks. 

As illustrated in \secref{sec: temporal}, each degree of active time shares the same embedding vector $\mathbf{ta}_i$, representing to what extent the positive feedback is. We feed this vector into the attention computation as extra click-level 
feedback. Now, $\alpha_i$ in \eqnref{eq:alpha} is modified to:
\begin{equation}
    \alpha_i = W_0 \times \sigma (W_1 \times \mathbf{xc}_i + W_2 \times \mathbf{ta}_i + b_0),
\end{equation}
where $W_2 \in \mathbb{R}^{d_n \times d_t}$ is the projection parameter that map the active time embedding with $d_t$ dimension into another dimension space. The contextual session vector $\mathbf{xc_s}$ still follows
\eqnref{eq:final_repre} and final session vector $\mathbf{x_s}$ is combined with $\mathbf{xt_s}$ and $\mathbf{xc_s}$.

\subsection{Modeling Negative Feedback}
\label{sec:negative feedback}
The most straight-forward and widely adopted negative sampling strategy is the random sampling 
from a non-clicked set of items, or from a global buffer with the last $N$ 
clicks~\cite{gabriel2019contextual}. 
The major problem of randomly sampled items is that these items might be completely unrelated to 
the user, posing too little challenge for the model to learn. On the contrary, an informative item should be able to confuse the model whether it has discovered more complex hidden meaning of user interaction or not.

While reading news, a user scrolls up and down the news stream, 
and the articles that are exposed to the user collectively 
form an impression list $Imp_u$. We take unclicked articles in $Imp_u$ as more informative negative signals than other candidates~\cite{xie2020deep} and thus 
we should treat them differently when counting loss, which means we should penalize the similarity between $\mathbf{xc_s}$ and those strong negative samples more strictly. This idea is similar to utilizing grayscale data~\cite{lin2020world} and contrastive learning~\cite{saunshi2019theoretical}, where we both consider the different degrees of information carried from different items.

As we discussed before, since the impression list is not always explicitly available, 
we assume an article is more likely to be in $Imp_u$ if it was published nearby an article that has
been clicked by $u$. Specifically, we sort the candidate articles according to their publishing time, and keep the nearby articles with the window size 300 and sample items from this window.
We aim to minimize the cosine score between $\mathbf{xc_s}$ and the vector $\mathbf{xc}_j$ of negative sample $j$ when 
$j\in Ne_u$, where $Ne_u\subseteq Imp_u$ is the set of negative samples for session $S_u$, thus we add this constraint into the final loss: 

\begin{equation}
    \label{eq:loss}
    \begin{split}
        \mathcal{L}_2 = - \frac{1}{|S|} & \sum_{S_u \in S}\sum_{j=1}^N ( y_j^u \log(\hat{y_j}^u) + (1-y_j^u)\log(1-\hat{y_j}^u) \\
        & +  \lambda \mathbbm{1}(j \in Ne_u) \log(\sigma(1-\mathbf{xc}_j^T\mathbf{xc_s}))),
    \end{split}
\end{equation}
where $\mathbbm{1}(\cdot)$ returns 1 if the expression is true, $\lambda$ is 
the weighting parameter of loss from negative articles. We jointly optimize these two losses 
with Adam optimizer.

\section{Experiments}
\label{sec:experiment}
In this section, we conduct experiments on three real-world news datasets: 
Adressa~\cite{gulla_adressa_2017}, Globo~\cite{gabriel2019contextual,moreira_news_2018} and MIND~\cite{wu2020mind}. 

\subsection{Experimental Setup}
\subsubsection{Data Preprocessing}
In dataset preprocessing, we treat a series of click events from one anonymous user 
within 30 minutes as a session. 
We discard the sessions no longer than 1 click.
To augment limited training data, for a session of $n$ clicks, we create $n-1$ mini-sessions, 
each starting from the first click of the original session and ending at the 2nd, 3rd through
the last click of the original session. The article clicked at the end of every mini-session is 
the label to be predicted. The dataset statistics are 
in \tabref{tb:dataset}, where each session is quite short. The public Globo dataset covers 16 days, and only provides the extracted vectors of articles. We only choose a subset of days (20 days)
in Adressa (the whole dataset lasts for 3 months) following~\cite{gabriel2019contextual} 
for simplicity. As for MIND, the data from the first several weeks is 
users' historical logs without the session information, so we choose the data from the last week (7 days). The active time is missing in MIND, but it explicitly contains the impression list of each session.

\begin{table}[th]
  \caption{Dataset statistics (after preprocessing)}
  \label{tb:dataset}
  \centering
  \begin{tabular}{l|c|c|c|c|c}
    \toprule
     Dataset & \# sessions & \# articles & \# topics & \tabincell{c}{clicks/\\session}  & \tabincell{c}{clicks/\\article} \\
    \midrule
    Globo & 1M & 45k & 461 & 2.69 & 64  \\
    Adressa & 0.5M & 12k & 23 & 2.79 & 117 \\
    MIND & 0.2M & 7k & 16 & 2.38 & 59 \\
    \bottomrule
  \end{tabular}
\end{table}

\subsubsection{Train/test set split}
In order to simulate the real-time recommendation 
scenario, we choose a standard framework~\cite{jugovac_streamingrec:_2018} 
which trains the model in the first few days and test the trained model in the remaining
days. Each dataset can be split into several folds, 
we will average the results over these folds.
For Globo dataset, we split every 4 days into one fold, with 3 days for training and 1 day for testing, and 4 folds in total. 
For Adressa dataset, we split every 10 days into one fold due to its fewer session data in one day, and we need to extend the training days to keep the similar size of training data with Globo. 
We average the metrics performance of each fold in the end. For MIND dataset, we leave the last day as the test set to make one fold. 
After data preprocessing, the ratio between training data size and test data size 
is around 6:1 for Globo, and 10:1 for the other two datasets. 


\subsubsection{Metrics}
During the test, given the first few click events in a session, the model generates 
a top-$k$ recommendation list $R$ with descending probabilities. 
We use widely-used metrics HR@$k$, NDCG@$k$ to 
measure the model's prediction accuracy.

Intra List Diversity (ILD@$k$)~\cite{symeonidis2020session} evaluates the topical/semantic 
diversity in $R$, and reflects the model's ability to recommend different items to the same user. 
\begin{equation}
  ILD@k = \frac{1}{|R|(|R|-1)}\sum_{a\in R}\sum_{b\in R}d(a,b),
\end{equation}
where $d(a, b)$ is a distance measure between item $a$ and $b$, and 
$d(a, b) = 1$ if item $a, b$ belong to different topics (categories), 0 otherwise.

Besides, we expect the system to recommend unseen items to surprise users. The content-based unexpectedness metric (unEXP)~\cite{kaminskas2014measuring} can be used to measure this kind of unexpectedness, which is calculated as follows:

\begin{equation}
  unEXP_u@k = \frac{1}{|R|}\sum_{a\in R}\frac{1}{|S_u|}\sum_{b\in S_u}d(a,b).
\end{equation}

\begin{table*}[h]
  \setlength{\tabcolsep}{3.9pt}
  \makebox[\textwidth][c]{
  \begin{threeparttable}[b]
  \caption{Main and ablation results ($k=20$ by default in our all tables). All results are averaged over all folds. The best baseline result on 
  each metric is marked with $*$ and overall best results are bolded. The ``Ours'' is our whole model and (-) means to ablate the corresponding module, where ``neut'', ``pos'' and ``neg'' respectively refer to our neutral, positive and negative feedback modules. The last column is to replace our negative sampling strategy with random sampling. 
    $\downarrow$ indicates performance drop over the whole model.}
    \label{performance-table}
    \centering
    
      \begin{tabular}{c|c|ccccccc|c|cccc}
      \toprule
      Datasets&Metrics&CBCF&STAN&GRU4Rec&SASRec&SRGNN&SGNNHN&STAMP&Ours&(-)neut&(-)pos&(-)neg&random \\ 
      \midrule
      \multirow{5}{*}{Adressa} & HR & 0.0957 & 0.1130 & 0.1120 & 0.1205 & 0.1152 & 0.1285 & 0.1287$^*$ & \textbf{0.1658} & 0.1344$\downarrow$ & 0.1619$\downarrow$ & 0.1658 &0.1646$\downarrow$\\ 
      \cline{2-14}
      & NDCG & 0.0341 & 0.0500 & 0.0511 & 0.0509 & 0.0536 & 0.0562 & 0.0575$^*$ & \textbf{0.0730}& 0.0613$\downarrow$ & 0.0690$\downarrow$ & 0.0720$\downarrow$ &0.0693$\downarrow$ \\ 
      \cline{2-14}
      & ILD & 0.2337 & 0.2409 & 0.8170 & 0.7856 & \textbf{0.8611}$^*$ & 0.8403 & 0.8445 & 0.8085& 0.8204 & 0.8249 & 0.8237 &0.8234\\ 
      \cline{2-14}
      & unEXP & 0.2509 & 0.2407 & 0.6949 & 0.8010  & 0.4754 & 0.8059$^*$ & 0.5728 & \textbf{0.8279}& 0.8243$\downarrow$ & 0.8333 & 0.8267$\downarrow$ &0.8346 \\ 
      \midrule
      \multirow{5}{*}{Globo} & HR & 0.1185 & 0.1273 & 0.1280 & 0.1409 & 0.1280 & 0.1414 & 0.1435$^*$ & \textbf{0.1852} & 0.1460$\downarrow$ & 0.1817$\downarrow$ & 0.1821$\downarrow$ &0.1847$\downarrow$\\ 
      \cline{2-14}
      & NDCG & 0.0474 & 0.0647 & 0.0599 & 0.0620 & 0.0627 & 0.0611 & 0.0698$^*$ & \textbf{0.0936}& 0.0727$\downarrow$ & 0.0907$\downarrow$ & 0.0940 &0.0933$\downarrow$\\ 
      \cline{2-14}
      & ILD & 0.3874 & 0.3087 & 0.9377 & \textbf{0.9864}$^*$ & 0.9248 & 0.9415 & 0.7980 & 0.8702 & 0.8362$\downarrow$ & 0.8685$\downarrow$ & 0.8927 &0.8739\\
      \cline{2-14}
      & unEXP & 0.3730 & 0.2921 & \textbf{0.9771}$^*$ &0.9690   & 0.6383  & 0.9467 &  0.8437 & 0.8358 & 0.8142$\downarrow$ & 0.8252$\downarrow$ & 0.8489& 0.8317$\downarrow$\\ 
      \midrule
      \multirow{5}{*}{MIND} & HR & 0.0315 & 0.0312 & 0.0338 & 0.0355 & 0.0334 & 0.0366 & 0.0371$^*$ & \textbf{0.0495} & 0.0445$\downarrow$ & - & 0.0471$\downarrow$ &0.0457$\downarrow$\\ 
      \cline{2-14}
      & NDCG & 0.0110 & 0.0142 & 0.0132 & 0.0139 & 0.0144 & 0.0122 & 0.0151$^*$ & \textbf{0.0211} & 0.0198$\downarrow$ & - & 0.0180$\downarrow$ &0.0204$\downarrow$\\
      \cline{2-14} 
      & ILD & 0.7166 & 0.3193 & 0.8662 & 0.8562 & 0.8706 & 0.8775$^*$ & 0.8452 & \textbf{0.8813} & 0.8779$\downarrow$ & - & 0.8808$\downarrow$ &0.8858\\ 
      \cline{2-14}
      & unEXP & 0.6039 & 0.1064 & 0.8578 & \textbf{0.8654}$^*$ & 0.4508 & 0.4514 & 0.7544 & 0.8617 & 0.8415$\downarrow$ & - & 0.8623&0.8680\\ 
      \bottomrule
    \end{tabular}
  \begin{tablenotes}
    \item[1] We conduct t significance test between the best score (if ours) and the second-best score for the main results, and the improvement is strongly significant as $p<0.001$. Between the results of the whole model and the ablation model, the decline is significant as $p<0.01$.
  \end{tablenotes}
  \end{threeparttable}
    }
\end{table*}

\begin{figure*}
  \begin{subfigure}[b]{0.8\columnwidth}
  \centering
  \includegraphics[width=\columnwidth]{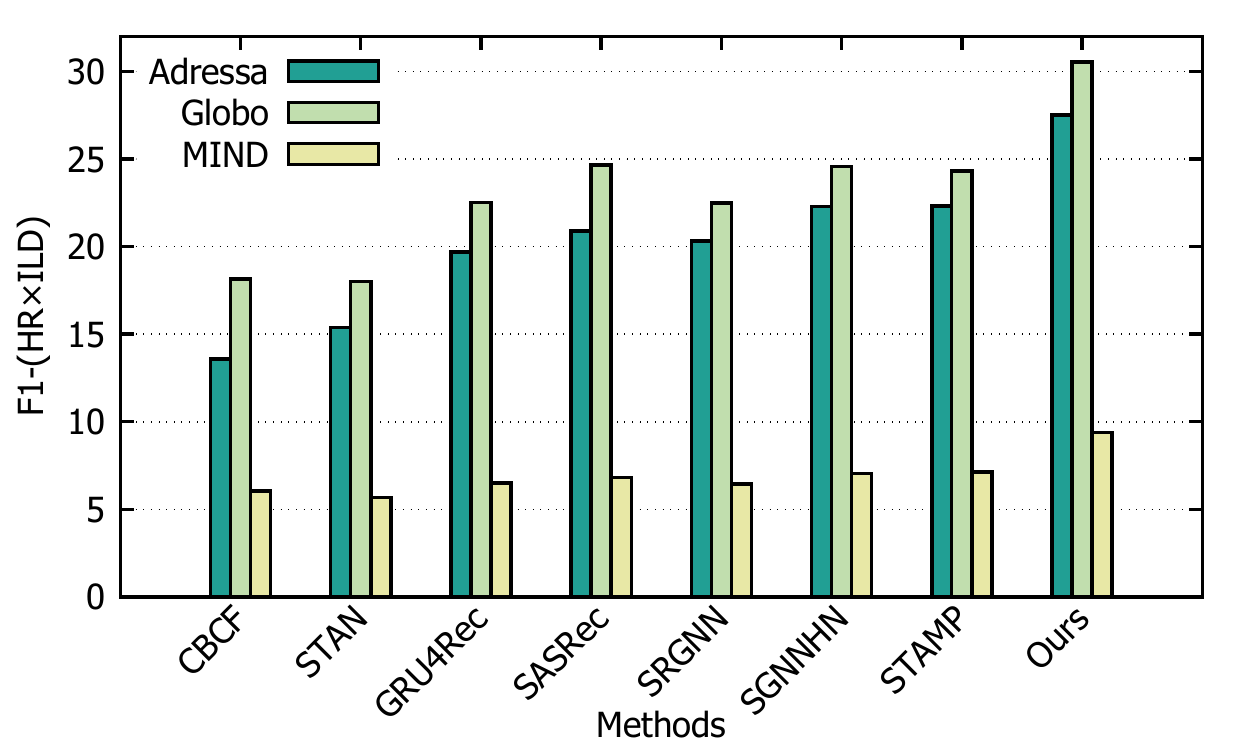}
  \caption{F1-(HR$\times$ILD)}
  \label{fig:hr}
  \end{subfigure}
  \hspace*{5mm}
  \begin{subfigure}[b]{0.8\columnwidth}
  \centering
  \includegraphics[width=\columnwidth]{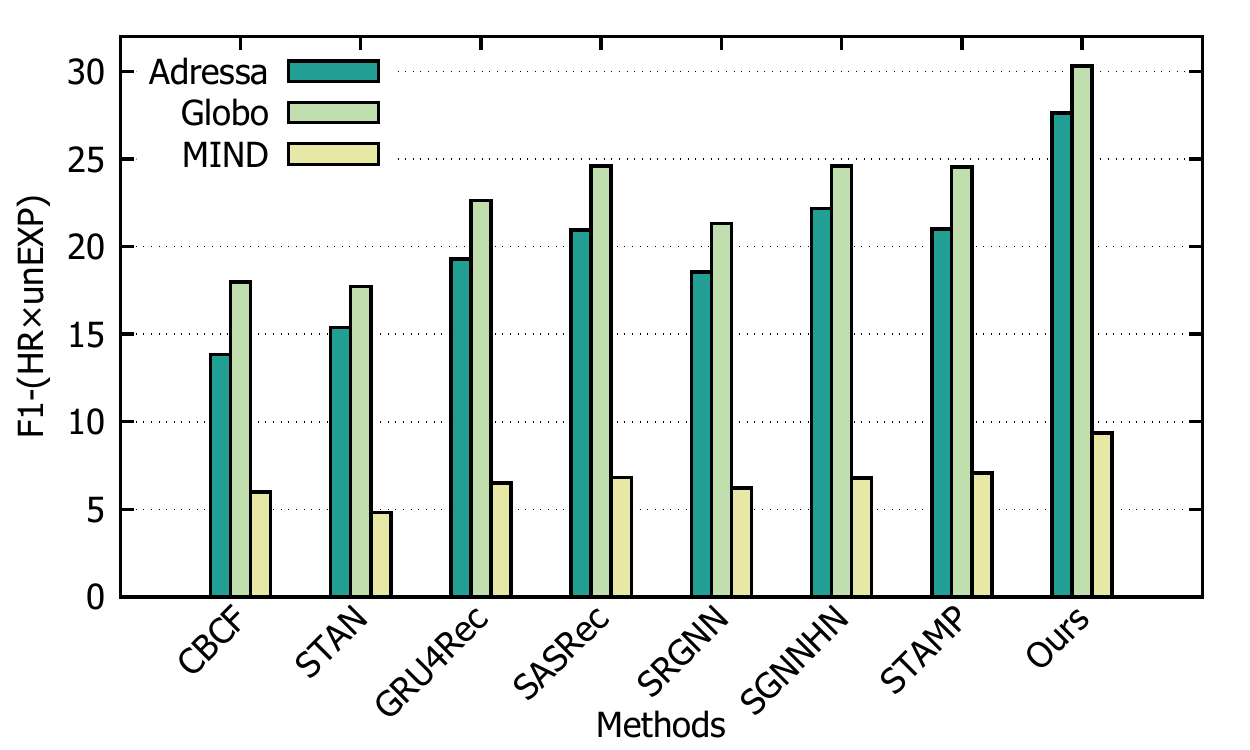}
  \caption{F1-(HR$\times$unEXP)}
  \label{fig:ndcg}
  \end{subfigure}
  \hfill
  \begin{subfigure}[b]{0.8\columnwidth}
  \centering
  \includegraphics[width=\columnwidth]{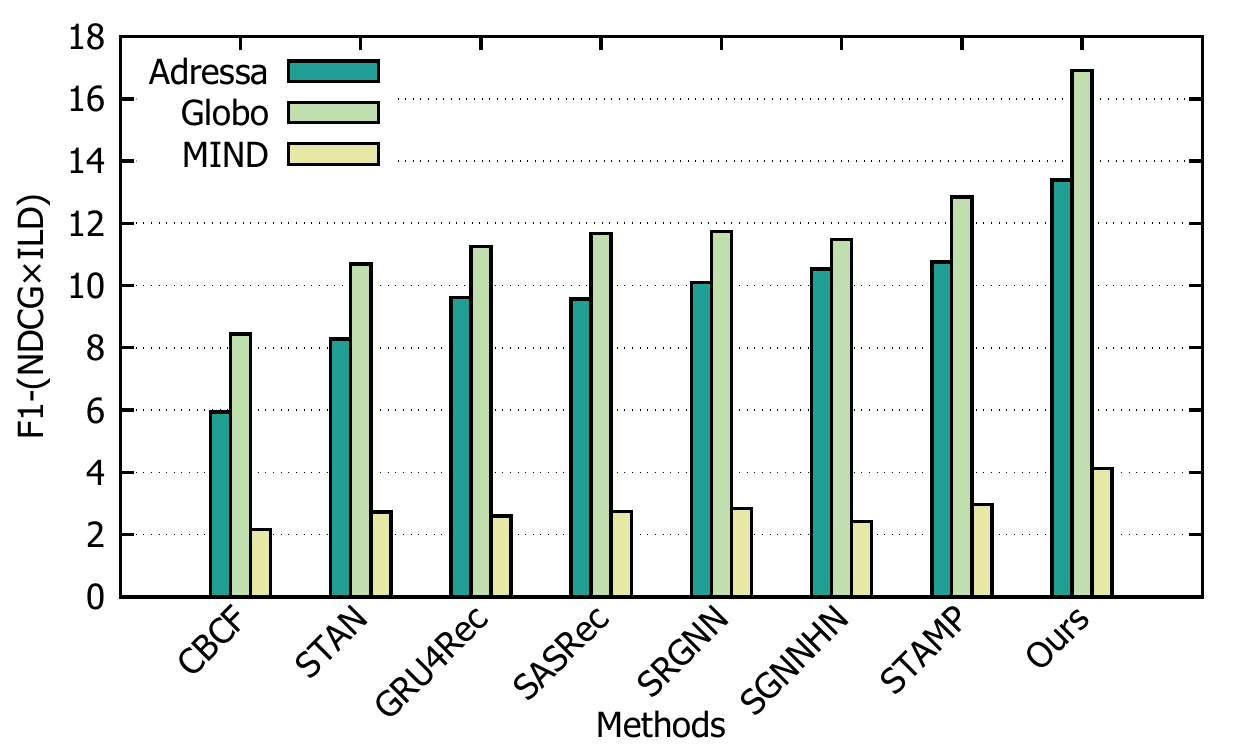}
  \caption{F1-(NDCG$\times$ILD)}
  \label{fig:ild}
  \end{subfigure}
  \hspace*{5mm}
  \begin{subfigure}[b]{0.8\columnwidth}
  \centering
  \includegraphics[width=\columnwidth]{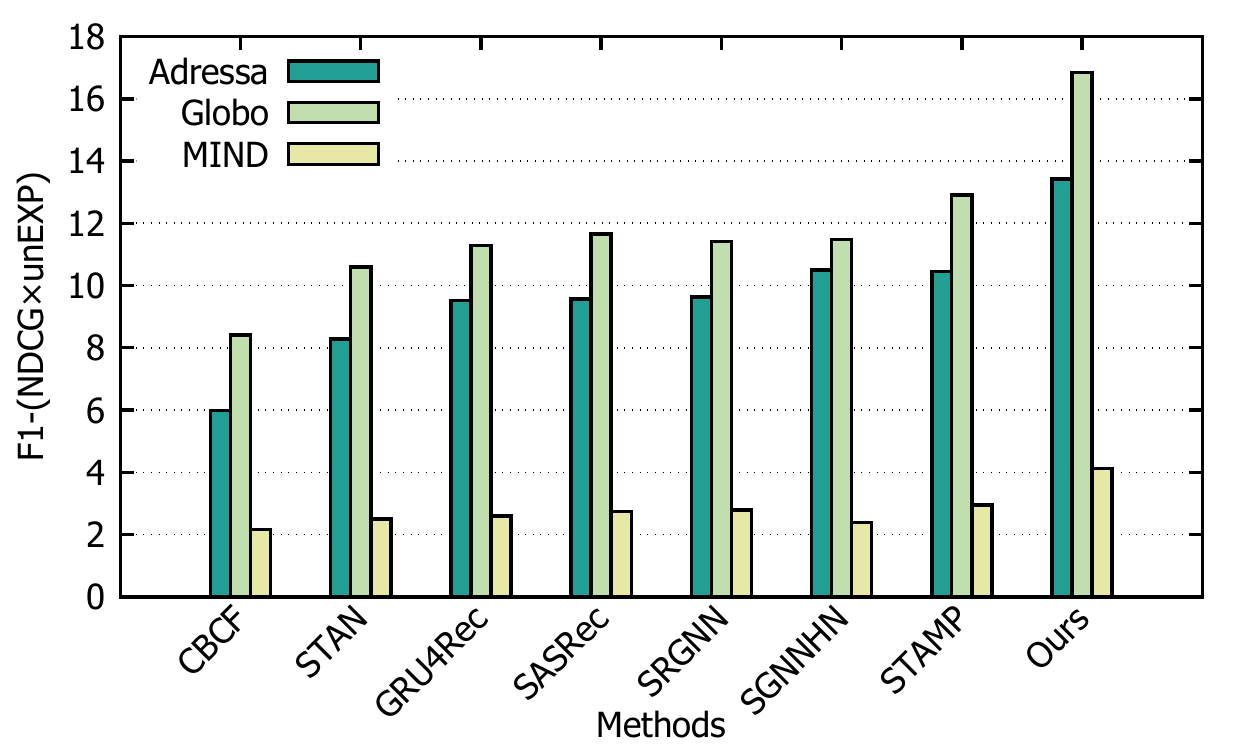}
  \caption{F1-(NDCG$\times$unEXP)}
  \label{fig:unexp}
  \end{subfigure}
  \caption{The graphical comparison of all methods on 4 different metrics and 3 different datasets.``Ours'' is our approach.}
  \label{fig:all}
\end{figure*}

\subsubsection{Baseline Algorithms}
Strong baselines are listed as follows. Their detail explanations are in \secref{sec:related session}.

\textit{Simple session-based recommenders:} Despite simplicity of some of those methods, they still have competitive accuracy. 
\textit{CBCF}~\cite{sottocornola2018session} is a news recommender combines Content-Based similarity with session-based Collaborative Filtering similarity.
\textit{STAN}~\cite{garg2019sequence} is an extended version of SKNN (Session-KNN) with three controllable temporal decay factors.

\textit{Session-based neural recommenders:}
  \textit{GRU4Rec}~\cite{hidasi2015session,hidasi2018recurrent} is a Gated Recurrent Unit for recommendation, building with gated recurrent neural networks, which is similar to LSTM in ~\cite{gabriel2019contextual}.
  \textit{SASRec}~\cite{kang_self-attentive_2018} is a self-attention based Sequential model, adopting Transformer architecture to model user's action.
  \textit{STAMP}~\cite{liu2018stamp} is a Short-Term Attention/Memory Priority Model introducing the attention mechanism to model the relationship of each historical click and the last click.
  \textit{SRGNN}~\cite{wu2019session} is a Session-based Recommender using Graph Neural Networks to capture complex transitions of items.
  \textit{SGNNHN}~\cite{pan2020star} is improved SR-GNN using Star Graph Neural Network.

  For session-based neural recommenders, we initialize the item embeddings with items' content vector for fair comparation.

\textit{Neural news recommendation approaches:}
\textit{CPRS}~\cite{wu2020CPRS} is a typical news recommendation approach that utilizes the textual feature of articles to model user's interests. It also uses the dwell time (i.e. active time) to measure user's satisfaction. We make this approach adapt to the session-based scenario. Since only Adressa dataset contains complete information for CPRS, we only compare it with our method in Adressa dataset and this is discussed in \tabref{tb:CPRS}.

\subsubsection{Implementation Details}
For fair comparisons, we apply all baselines and our method to the same augmented data and train 
models on one GTX1080Ti GPU\footnote{The implementation of our approach and baselines is released at \href{https://github.com/summmeer/session-based-news-recommendation}{https://github.com/summmeer/session-based-news-recommendation}}.
We use the same latent dimension $d_n=d_c=250, d_t=64$, choose different learning rate in $\{0.002, 0.001, 0.0005\}$, batch size in $\{512, 1024, 2048\}$ and other hyper-parameters to select the best model using early stopping strategy based on the HR@20 score on the validation set, which is the last 10\% of samples in the training set sorted by time. All embedding tables are normally initialized with 0 mean, 0.002 standard deviation, and for weighting parameters 0 mean, 0.05 standard deviation.

\subsection{Main Results}
\label{sec:mainres}

In \tabref{performance-table} and \figref{fig:all}, we compare the performance of all baselines and our
approach, and we can make the following observations.

Non-neural methods CBCF and STAN are either considering the content 
information or the recency of the current session, and their results are somehow 
comparable to deep learning methods in three datasets. However, they generate recommendation lists with low diversity/novelty, mainly because their simple algorithms cannot capture enough personalized information. For session-based approaches, generally speaking, STAMP and SGNNHN yield better performance on HR and NDCG, but not always good at ILD/unEXP, while SASRec and SRGNN recommend more diverse but less accurate, showing the trade-off between diversity and accuracy. From the user's aspect, though, when ILD/unEXP is over a threshold (like around 0.8\footnote{We recruit two volunteers to measure the diversity of 50 samples to get this consensus}), it's hard for them to distinguish the difference, thus the ILD/unEXP score of our model is bearable. 

As for our whole model, when compared with STAMP, it performs better or close on both accuracy and diversity. This result shows that \textbf{our model mitigates the dilemma between accuracy and diversity} 
to a certain extent. In the MIND dataset, 
the improvement is comparatively small and the possible reasons are: 
on the one hand, MIND did not provide active interval, nor did they give click time of each article (just start time of a session), 
we cannot get positive feedback from the data; on the other hand, from the results of CBCF, 
we assume the article transition information is too sparse and thus it is hard to recommend. 
Note that this dataset is not designed for the session-based recommendation, hence some information 
may be inaccurate (e.g., one session may last for days, longer than 30 minutes).

\subsection{Effectiveness of Components}
From \tabref{performance-table} we can verify the effects of modeling user positive negative and neutral implicit feedback in our model. Compared with the whole model, there is a huge drop after removing neutral information and this is the most consistent over all metrics, which reveals the importance of neutral information (temporal information), and we will discuss the modeling of it in detail (\secref{sec:t}). We cannot get positive feedback from MIND so this column is empty, and the reasons why the improvement is limited for MIND are analyzed previously.

Adressa provides \textbf{the most complete information}, which not only releases the original 
text of articles instead of the extracted vectors in Globo, but also gives the accurate active time 
of the user in each article, while we can only estimate the active time by the interval between 
two successive clicks for Globo, which may not be accurate. After removing the positive implicit feedback module, in Adressa dataset, the HR and NDCG drop by 2.4\% and 5.5\% respectively, while in Globo dataset, they drop by 1.9\% and 3.1\%. The positive information performs similarly in both Adressa and Globo datasets, implying that our approximate estimation is reasonable. Further, the positive implicit feedback is more favorable on the Adressa dataset due to the more precise information.

We observe that negative information is less effective than positive information, especially by diversity/novelty metrics. 
One explanation is that the negative samples from the impression list are reconstructed based on their publishing time, so the information is not totally reliable. Negative sampling module lowers diversity, possibly because in the dataset the negative samples and the positive article usually belong to different categories, thus adding this module forces the model to recommend similar articles to the positive one. Negative feedback is better modeled in MIND due to its complete impression data. To verify the effect of the negative sampling strategy more accurately, we set the control group with random sampling, and we find that even though the random sampling would decrease the performance slightly, our negative feedback shows superior performance over it. The possible reason for the worse performance of using random sampling is that randomly sampled negative items have the possibility to be liked by this user, and this module imports some noise instead because this sample strategy does not consider \textbf{what the user really likes}.

\subsection{Effects of Positive Feedback}
CPRS considers the active time to represent the user's satisfaction using personalized reading speed, which is quite similar but more complex than our positive feedback modeling. We firstly modify this method to meet the 
session-based scenario, and secondly plug the personalized reading speed into our model. 
The experiment is conducted in Adressa due to its complete information 
(Globo does not provide the text-level content and the active time is missing for MIND). 
In \tabref{tb:CPRS}, the poor score of CPRS shows that when it is adapted to the session-based scenario, 
limited interactions are the bottleneck. When we adopt their personalized reading speed instead of 
the reading time, there is no significant improvement, and we hypothesize that for this dataset, 
when news reading \textbf{the reading speed is quite similar} for different users.

\begin{table}[th]
  \caption{Results of CPRS and CPRS module plugged into our model in Adressa dataset.}
  \label{tb:CPRS}
  \centering
  \begin{tabular}{c|cccc}
    \toprule
    Methods  & HR & NDCG & ILD & unEXP\\
    \midrule
    Our whole model & 0.1658 & 0.0730 & 0.8085 & 0.8279\\
    \midrule
    CPRS & 0.0812 & 0.0371 & 0.8191 & 0.8109\\
    Ours using speed & 0.1641 & 0.0674 & 0.8457 & 0.8245\\
    Ours using 1-d $t_i$ & 0.1603 & 0.0705 & 0.8293 & 0.8220\\
    \bottomrule
  \end{tabular}
\end{table}

To verify the effectiveness of the duration encoder, we compare it with the continuous active time vector $t_i$ regarded as a one-dimensional vector instead of using distinct categories. As the result shows, the accuracy score of 1-d $t_i$ is inferior to our whole model, indicating that our duration encoder catch more personalized information by bucketizing it.

\subsection{Effects of Negative Feedback}
In this section, we first validate our assumption for the negative user feedback,
which is that articles whose publishing time is close to the clicked articles are
likely presented to the user, or within their impressions.
We do that on the MIND dataset, in which the real impressions and the
publishing time of articles are all available for the sessions. 
For each session $S_u$, we sample negative items $Ne_u$ using our strategy, and compute Jaccard similarity between $Ne_u$ and real $Imp_u$, the overall score is 0.0062 when $|Ne_u=100|$, compared with 0.0044 when random sampling.
This shows that our assumption is reasonable and our strategy can \textbf{better reconstruct the impression list}.

\begin{figure}[th]
  \begin{subfigure}{0.49\columnwidth}
  \centering
  \includegraphics[width=\columnwidth]{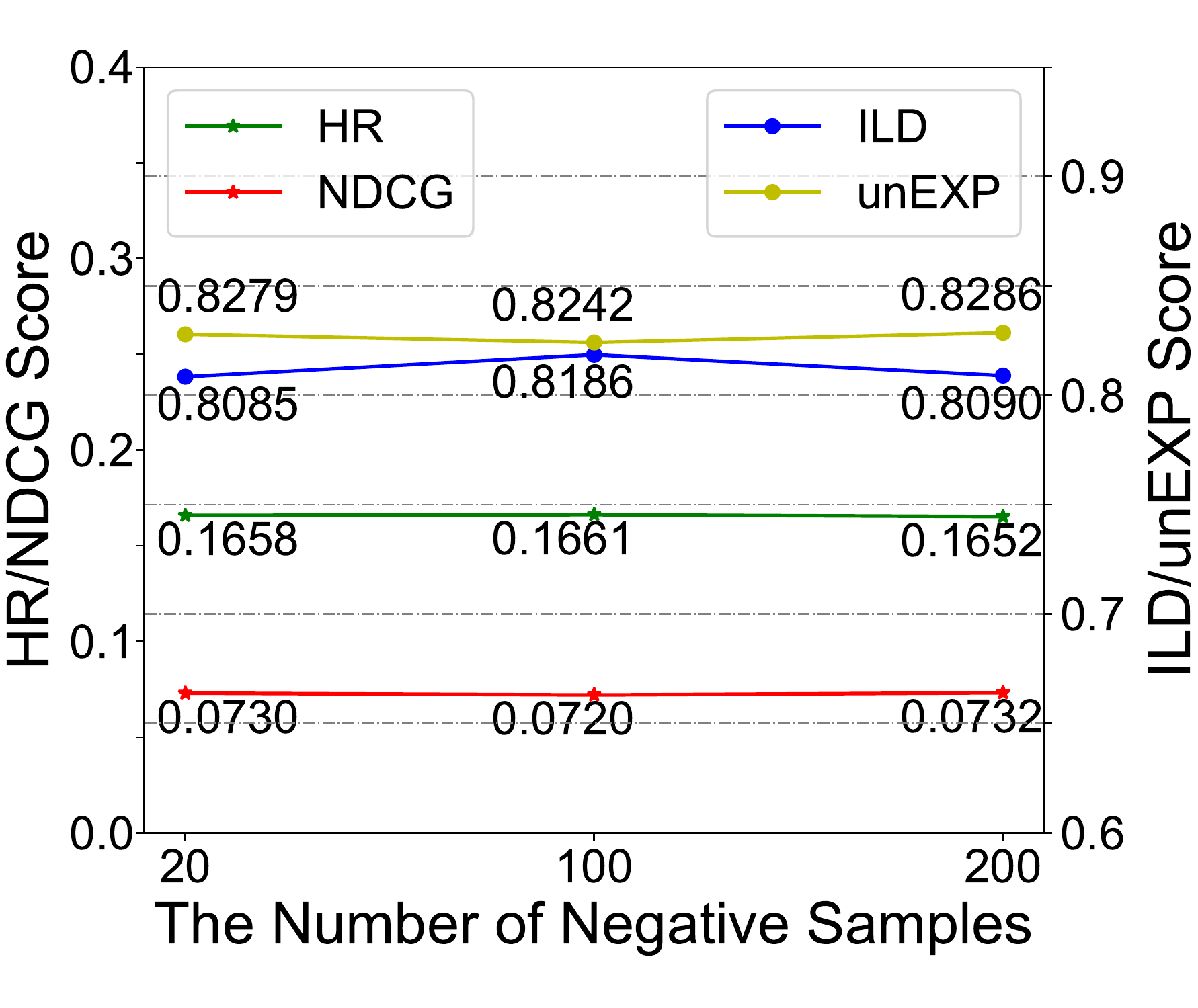}
  \caption{Results for different $|Ne|$}
  \label{fig:para_a}
  \end{subfigure}
  \begin{subfigure}{0.49\columnwidth}
  \centering
  \includegraphics[width=\columnwidth]{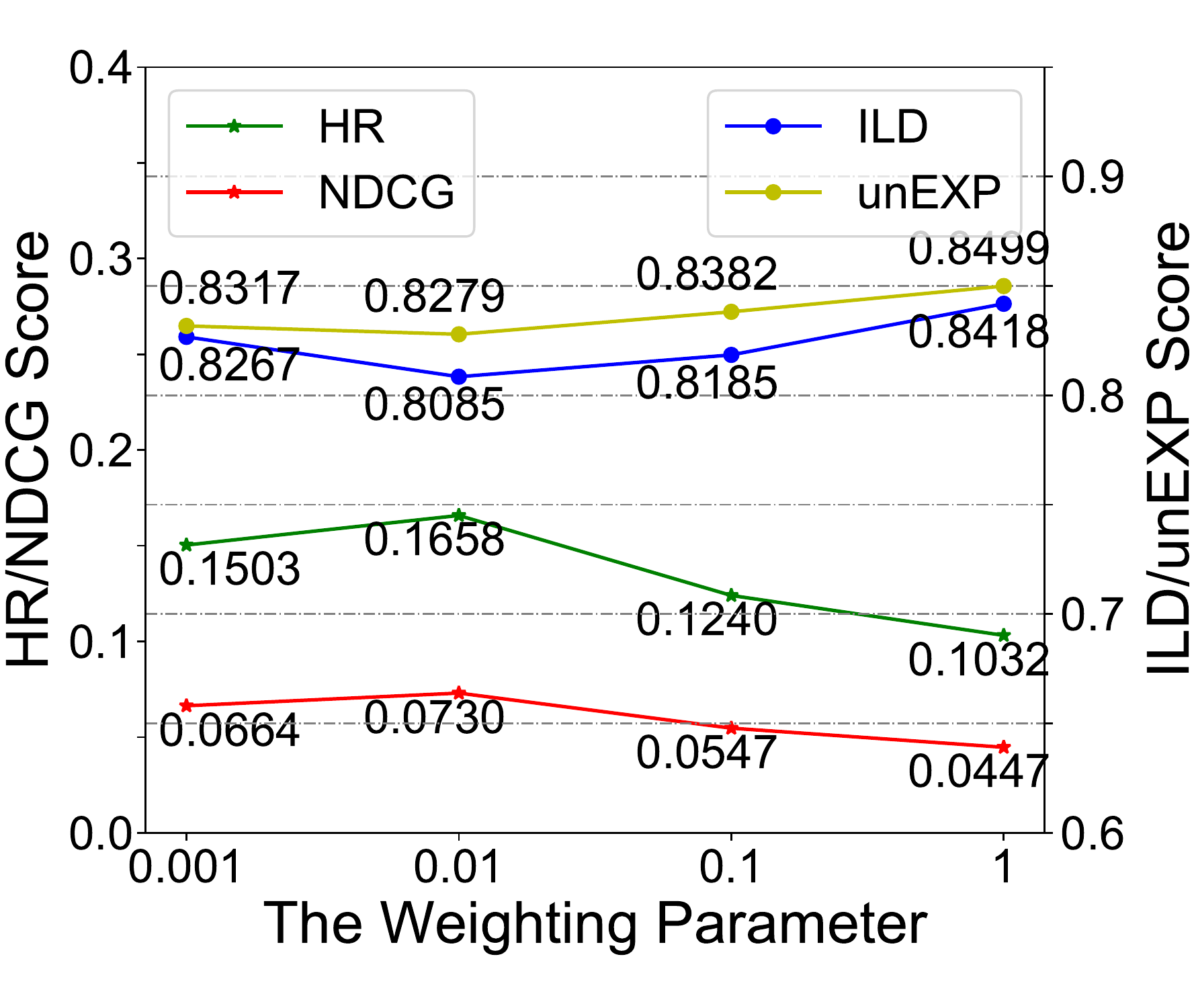}
  \caption{Results for different $\lambda$}
  \label{fig:para_b}
  \end{subfigure}
  \caption{Hyper parameter discussion.}
\end{figure}

We further conduct a parameter analysis on loss weights $\lambda$ and the number of negative samples $|Ne|$ 
in \eqnref{eq:loss}. We report results on Adressa as an example, and results from other datasets are similar. 
The number of negative samples does not matter much, as shown in \figref{fig:para_a}, 
so for simplicity we choose $|Ne|=20$. The performance gets worse if $\lambda$ is set too low or too 
high in \figref{fig:para_b}, we conclude that the negative loss is useful but too many weights on it will \textbf{harm} the learning of the user's positive feedback.

\subsection{Temporal Representations}
\label{sec:t}
Since both the start time and the publishing time use the temporal encoder, 
we wonder if it would be better to have them share the embedding space. 
\tabref{tb:temporal} shows the findings. 

\begin{table}[th]
  \caption{Different ways of utilizing temporal information in Adressa, where ``p'' stands for the publishing time and ``s'' stands for the start time.}
  \label{tb:temporal}
  \centering
  \begin{tabular}{c|cccc}
    \toprule
    Methods  & HR & NDCG & ILD & unEXP \\
    \midrule
    whole (shared) & 0.1658 & 0.0730 & 0.8085 & 0.8279\\
    whole (no share) & 0.1620 & 0.0727 & 0.8310 & 0.8215\\
    \midrule
    whole-p-s & 0.1353 & 0.0612 & 0.8280 & 0.8276\\
    whole-p & 0.1344 & 0.0613 & 0.8204 & 0.8243\\
    whole-s & 0.1620 & 0.0726 & 0.8325 & 0.8415\\
    \bottomrule
  \end{tabular}
\end{table}

We can see that sharing the time embedding between publishing time and
session start time has clear advantages in most of the metrics. This is
because publishing time is associated with every article and there are a lot of
such data for training, whereas the session start time suffers from lack
of data and is less trained. Training the two jointly implicitly helps each 
other. It also makes physical sense because a Monday is a Monday regardless
a story breaks on that day or a reader pops in to read that story on that day.

The ablation tests of using only publishing time or only start time in
\tabref{tb:temporal} also clearly indicates that temporal modeling both from
the item point of view and the user point of view are useful in garner latent
information between the two. In other words, \tabref{tb:temporal} shows that
the neutral feedback works.

\begin{figure}[th]
  \centering
  \includegraphics[width=\columnwidth]{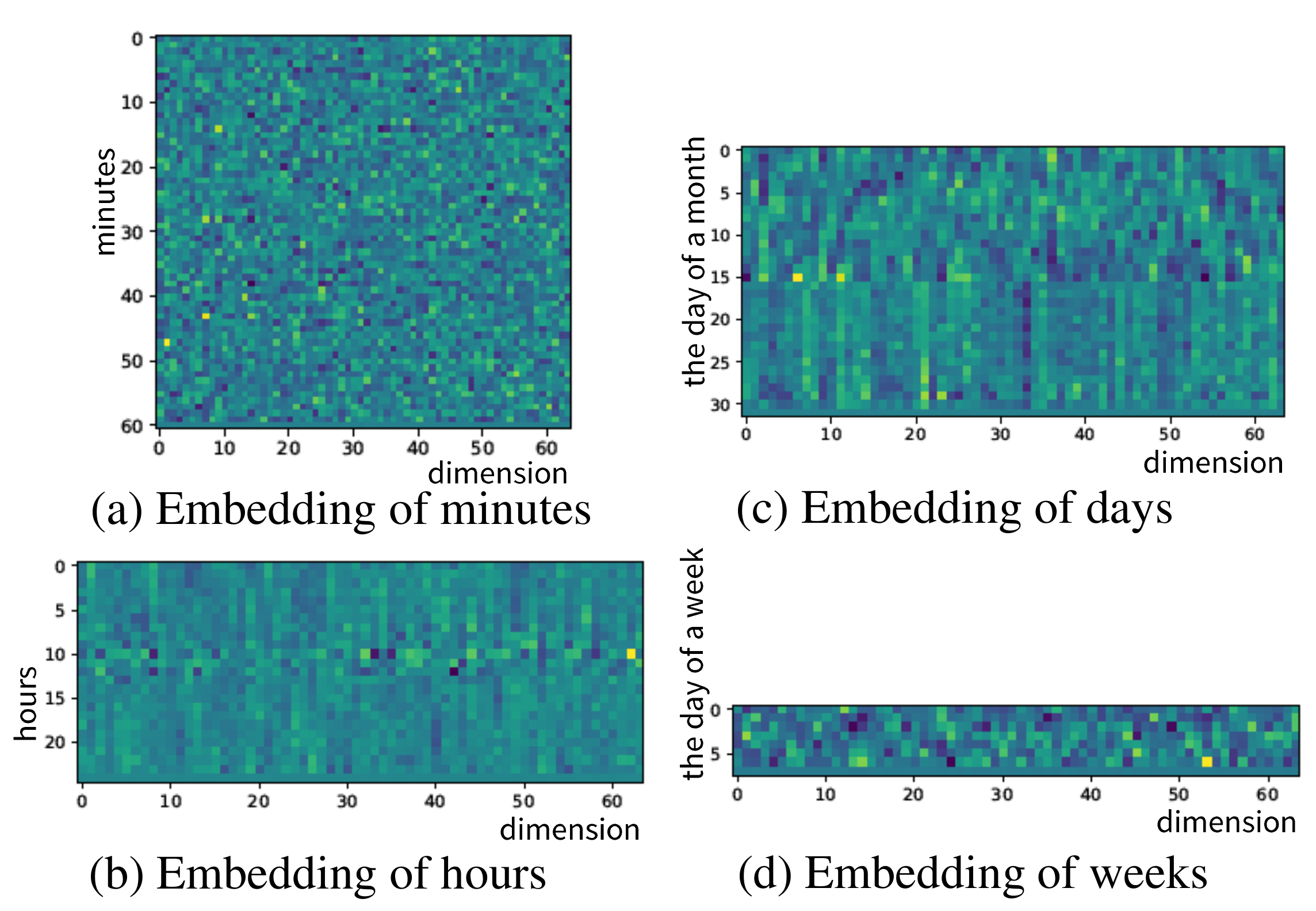}  
  \caption{The visualization of time embedding tables for the day of a month, the day of a week, hour and minute, trained on Globo dataset.}
  \label{fig:temporal}
\end{figure}

To give some concrete evidence that the time embedding that we train
carries some physical meanings, we visualize the embedding tables for
\textit{minutes}, \textit{hours}, \textit{weekdays} and \textit{days
of a month}, in \figref{fig:temporal}, which is trained on a subset of the Globo dataset.
Some interesting patterns can be observed. For example, the representation
of the minutes is rather uniform and random, because news publishing and
reading can happen any minute of an hour. But there are certainly more
activities at certain hours during a day. There are also
some irregular patterns for weekends as shown in (c).
Finally, because we only have the first 15 days of training data in this dataset,
values for dates 16-31 in (d) are not fully trained, which only has the chance to update when the publishing date of articles falls in the range of 16-31, but most of articles are published nearby the click time according to the dataset statistics.

\subsection{Discussion}
\label{sec:discuss}

\subsubsection{Article cold-start}
\label{sec:itemcold}
For news recommendation, all methods suffer from article cold-start problem due to the continuously published news, the analyses of the article cold-start scenario can help us 
figure out where our improvement comes from. Another concern about the article cold-start problem is that if fresh articles can not get exposure reasonably, they will suffer from the Matthew effect and will not be clicked anymore. According to~\cite{gabriel2019contextual}, instead of using user-oriented metric ILD/unEXP, we thus consider the system-oriented Item Coverage (COV@$k$) as an additional metric. COV@$k$ is also called ``aggregate diversity'', and reflects the model's ability to recommend different items to different users, which forces the model to make a larger fraction of the item catalog visible to the users. We compute COV@$k$ as the number of distinct articles that appeared in any $R$ divided by the number of distinct last clicked articles in the test set. 
\begin{table}[th]
\caption{Cold-start performance on Globo in the first fold. HR score is  listed as percentage due to its relatively small value, and all scores are reserved to the second decimal.}
\label{tb:cold-start}
\centering
\begin{tabular}{c|c|c|c|c|c|c}
  \toprule
  \multirow{2}{*}{Methods}  & \multicolumn{2}{c|}{Cold(80.3\%)} & \multicolumn{2}{c|}{non-Cold(19.7\%)} & \multicolumn{2}{c}{Total} \\ \cline{2-7} 
    & HR(\%) & COV & HR(\%) & COV & HR(\%) & COV   \\ 
  \midrule
  CBCF & 3.69   & 5.06  &  24.88 & 5.54 & 7.87  & 3.95 \\
  STAN & -  & - &  26.52  &  1.63 & 5.22   & 0.93  \\ 
  \midrule
  GRU4Rec & 1.51  & 0.03  & 20.93 & 0.88 &  5.33  & 0.50 \\
  SASRec & 0.80  & 0.01  & 23.35 & 1.28 & 5.25 & 0.73 \\
  SR-GNN & 1.00  & 0.01  & 23.65 & 0.99  & 5.46 & 0.57\\ 
  STAMP & 1.72 & 0.01 & 21.84  & 1.04 & 5.68 & 0.59 \\
  SGNNHN & 0.89 & 0.01 & 24.86 & 0.05 & 5.61 & 0.04 \\
  \midrule
  Ours & 4.96 & 0.74 & 25.27 & 1.87  & 8.96  & 1.20  \\
  \bottomrule
\end{tabular}
\end{table}

\tabref{tb:cold-start} lists the results of one fold in Globo dataset, and we choose it because it suffers from the most severe cold-start problem. 
For cold situation, where the test articles are completely disjoint from
the training data, STAN does worse because it can not handle unseen items.
Deep learning methods tend to predict the same articles for different users. 
Even though methods like SASRec yields not bad results, 
the models tend to overfit to popular articles. Our model, on the other hand, not only performs well on HR@20 but also gets the comparable COV@20 score, and the difference with the
other deep learning methods is remarkable. 
In non-cold situation, the performance of all methods is close. 
The overall recommendation results largely depend on how a method does for 
cold-start scenarios.

\subsubsection{User cold-start}
\label{sec:usercold}
\begin{figure}[!htp]
  \centering
  \includegraphics[width=0.98\columnwidth]{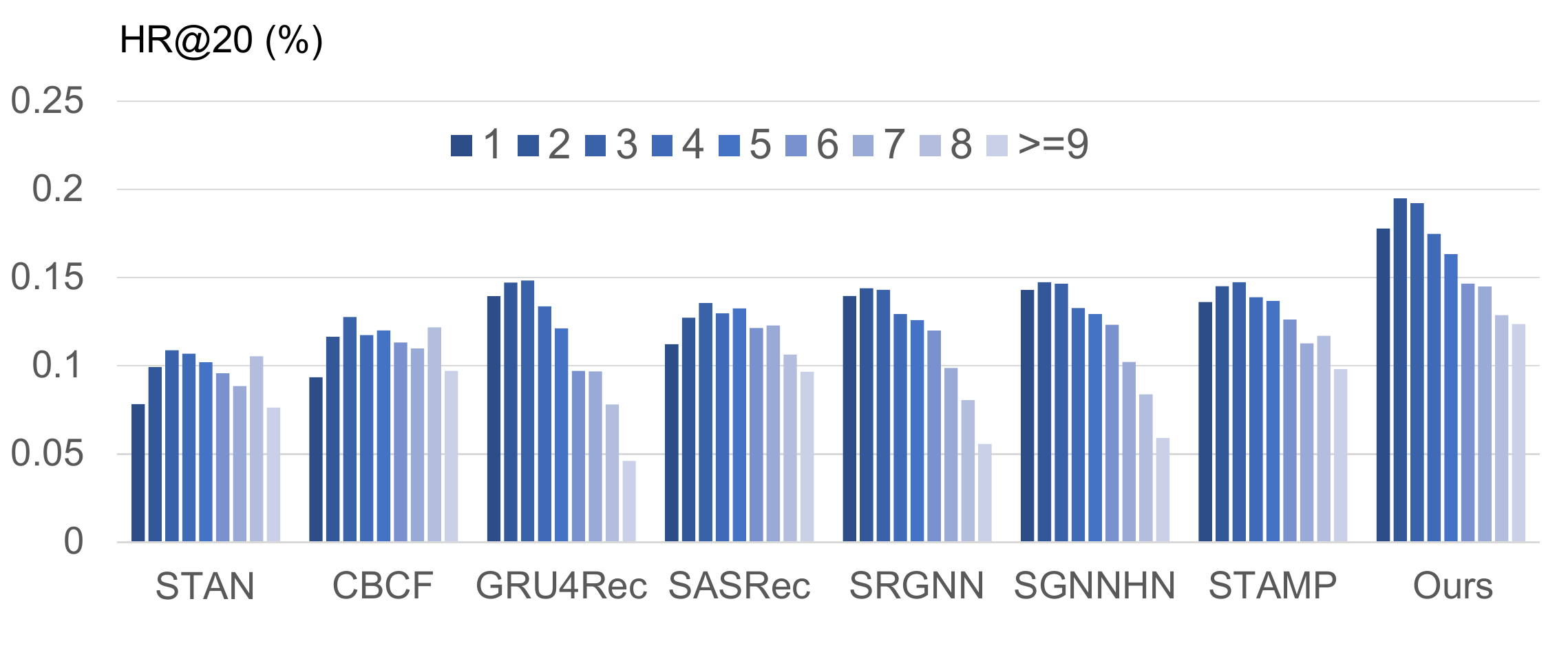}
  \caption{Accuracy with different session lengths.}
  \label{fig:inputlen}
\end{figure}

Since anonymous news sessions are short with the average length of less than 3, 
the user cold-start problem is severe. We show the results for different input lengths in 
\figref{fig:inputlen}. We report results on Globo, and other datasets perform similarly. Interestingly, our model reaches its peak accuracy from length 1 to 2. 
In contrast, other methods all reach the peak at 3.
This shows our model is capable of \textbf{capturing user interests earlier} in the session 
by leveraging the user's implicit feedback. For longer input length, 
the difference between our model and others narrows, 
indicating the similar ability to predict a user's interests given a longer history. We can observe that the performance drops with the longer length input, and this may contribute to the noise that is imported from the longer reading history, which means it is harder to recommend when considering the longer and more diverse interests from users.

\section{Related Work}
In this section, we discuss works in the area of news recommendation and session-based recommendation in news and other domains like e-commerce or music, and we also compare them with our proposed approach.

\subsection{News Recommendation}
First, the news recommendation task can be formulated as conventional recommendation tasks, the account of a user is reserved and articles are recommended based on users' long-term click history. Some works use a well-designed graph network to represent the target user and the clicked article~\cite{hu2020graph,ge2020graph}.
In this situations, the relation of items and users are well exploited. Unfortunately, in real-time scenarios, new articles and anonymous users emerge, causing a severe cold-start problem. Then if we want to capture users' preferences
within sessions and recommend articles with their several interactions as input, this kind of approach with the static user-item matrix is not suitable. Some propose incremental matrix factorization algorithm based on classic MF algorithm by adding a new user or a new article to the matrices with a random initialization~\cite{al2018adaptive}, and others apply meta-learning which aims to train a model that can rapidly adapt to a new task with a few examples~\cite{lee_melu:_2019}, but do not solve the problem fundamentally.

Second, some news recommendation systems use clicked articles to represent a user, which can be adaptive to anonymous users. Some of them encode the article text with fine-grained attention mechanism~\cite{zhu2019dan,wu_neural_2019-1,wu2019npa,wang2020fine}, some consider the relation between the dwell time of the user and satisfaction of the user~\cite{wu2020CPRS}, and others use the knowledge graph of entities in the news title as affiliated information~\cite{wang2018dkn,wang_ripplenet:_2018}. They mainly focus on the textual feature of articles in order to aggregate users' preference while paying less attention to the click behavior. Although they can be applied for anonymous users by replacing long-term history clicks with articles within the session when fetching user representations, challenges are that they cannot take full advantage of the textual information due to the limited interactions and the overload of training cannot be avoided. Besides, they evaluate their methods by classifying the real clicked article and several constructed distractors from the impression list, and this is not consistent with the real recommendation scenario, where the recommender retrieves top-K recommendation lists from all candidates.

For the rest of the work, one uses the information of how frequent user returns to help improve recommendation~\cite{zheng2018drn}, another work jointly models click, unclick and dislike as explicit/implicit feedback~\cite{xie2020deep}, and others excavate the quality of news articles~\cite{lu_quality_2019} or the backtracking behavior as the user feedback~\cite{smadja_understanding_2019}.
\subsection{Session-based News Recommendation}
\label{sec:related session}
Many online recommender systems are proposed to deal with the session-based scenarios~\cite{epure_recommending_2017,zhou_variational_2019}, where the user interaction information is limited and items are increasingly generated. Usually session-based news recommendation approaches integrate content-based similarity~\cite{sottocornola2018session}, and many of them introduce external knowledge to recommend top-K similar articles~\cite{symeonidis2021session,sheu2020context,sheu2021knowledge}. Some recommenders consider the past sessions of the same user~\cite{zhang2019dynamic,zhang2018deep}, which is not consistent with our anonymous settings, and that is why we do not compare experiment results with them.

Many other session-based recommenders are in the e-commerce domain, which can also be converted to deal with news articles. Here RNN, LSTM and GNN possess properties that make them attractive for sequence modeling of user sessions~\cite{guo_streaming_2019,hidasi2015session,wang2019modeling,gabriel2019contextual,wu2019session}. Further, a hybrid encoder with an attention mechanism is introduced to model the sequential behavior of users~\cite{li2017neural,liu2018stamp,xu2019time,song_islf_2019,zhang_feature-level_2019}. 
Besides, many sequential recommendation systems~\cite{pereira2019online,xu2019graph} on music listening, games playing construct assorted RNN-related architectures (e.g, RCNN~\cite{xu_recurrent_2019}, GRU~\cite{hidasi2018recurrent}, HGN~\cite{xiao2019hierarchical,ma2019hierarchical}), showing RNN's high capacity to modeling user shift preference.

Although above works naturally take the content information and preference shifting into account, the implicit user feedback are neglected. When sampling negative articles, an adaptive negative sampling method based on GAN is proposed~\cite{wang_neural_2018}. Beyond that, few works pay attention to the implicit meaning of negative samples. Randomly sampling from such continuously increasing and high-volume news articles might be fast but will not be effective enough.

\subsection{The Use of Temporal Information}
Sequence and Time-Aware Neighborhood (STAN)~\cite{garg2019sequence} takes vanilla SKNN as its special case. They build static time decay functions for three factors: the position of an item in the current session, recency of a past session w.r.t. to the current session, and the position of a recommendable item in a neighboring session. This approach can be regarded as rule-based SKNN, with exponential decay function, and the experiment result on e-commerce websites even outperforms some deep-learning-based approaches. 
In the deep learning model, some works design different temporal kernel functions or decay functions for different consumption scenarios~\cite{wang2020make,wu2020deja,zhang2019dynamic}. However, these functions of news articles is fixed, which may undermine the ability to model user's short-term preferences towards different articles. Dwell time is considered in~\cite{wu2020CPRS} as the user satisfaction, but the difference of users' reading speed is hard to capture in our session-based scenario. A time-interval-based GRU is proposed to model user session-level representations~\cite{lei_tissa_2019}, and some work~\cite{rakkappan2019context,xu2019time,wu_recommender_2019} treat the time feature of interactions as a temporal context, while they fail to consider the publishing/click/active time in the different dimension.

\section{Conclusion}
Session-based recommendations are indispensable under the streaming like 
or real-time scenario when users' historical records are unavailable. 
By leveraging the positive and negative implicit feedback from users, as well
as properly modeling the times in the problem, 
our proposed method is simple but effective to improve the 
trade-off between accuracy, diversity and surendipity, 
as shown in experimental results. For further investigation, our positive/negative modules can be plugged into other sophisticated 
session-based recommendation approaches; the automatic diversity metric may not always accord with the user experience, and attention can be paid towards the real user satisfaction; the temporal encoder can encode the physical meaning of the date-time, so maybe the pre-trained temporal embedding can improve the model's understanding of tasks which contain temporal information.

\bibliographystyle{ACM-Reference-Format}
\bibliography{references}


\begin{thebibliography}{61}


\ifx \showCODEN    \undefined \def \showCODEN     #1{\unskip}     \fi
\ifx \showDOI      \undefined \def \showDOI       #1{#1}\fi
\ifx \showISBNx    \undefined \def \showISBNx     #1{\unskip}     \fi
\ifx \showISBNxiii \undefined \def \showISBNxiii  #1{\unskip}     \fi
\ifx \showISSN     \undefined \def \showISSN      #1{\unskip}     \fi
\ifx \showLCCN     \undefined \def \showLCCN      #1{\unskip}     \fi
\ifx \shownote     \undefined \def \shownote      #1{#1}          \fi
\ifx \showarticletitle \undefined \def \showarticletitle #1{#1}   \fi
\ifx \showURL      \undefined \def \showURL       {\relax}        \fi
\providecommand\bibfield[2]{#2}
\providecommand\bibinfo[2]{#2}
\providecommand\natexlab[1]{#1}
\providecommand\showeprint[2][]{arXiv:#2}

\bibitem[\protect\citeauthoryear{Al-Ghossein, Murena, Abdessalem, Barr{\'e},
  and Cornu{\'e}jols}{Al-Ghossein et~al\mbox{.}}{2018}]%
        {al2018adaptive}
\bibfield{author}{\bibinfo{person}{Marie Al-Ghossein},
  \bibinfo{person}{Pierre-Alexandre Murena}, \bibinfo{person}{Talel
  Abdessalem}, \bibinfo{person}{Anthony Barr{\'e}}, {and}
  \bibinfo{person}{Antoine Cornu{\'e}jols}.} \bibinfo{year}{2018}\natexlab{}.
\newblock \showarticletitle{Adaptive collaborative topic modeling for online
  recommendation}. In \bibinfo{booktitle}{\emph{Proceedings of the 12th ACM
  Conference on Recommender Systems}}. \bibinfo{pages}{338--346}.
\newblock


\bibitem[\protect\citeauthoryear{Cheng, Koc, Harmsen, Shaked, Chandra, Aradhye,
  Anderson, Corrado, Chai, Ispir, et~al\mbox{.}}{Cheng et~al\mbox{.}}{2016}]%
        {cheng2016wide}
\bibfield{author}{\bibinfo{person}{Heng-Tze Cheng}, \bibinfo{person}{Levent
  Koc}, \bibinfo{person}{Jeremiah Harmsen}, \bibinfo{person}{Tal Shaked},
  \bibinfo{person}{Tushar Chandra}, \bibinfo{person}{Hrishi Aradhye},
  \bibinfo{person}{Glen Anderson}, \bibinfo{person}{Greg Corrado},
  \bibinfo{person}{Wei Chai}, \bibinfo{person}{Mustafa Ispir}, {et~al\mbox{.}}}
  \bibinfo{year}{2016}\natexlab{}.
\newblock \showarticletitle{Wide \& deep learning for recommender systems}. In
  \bibinfo{booktitle}{\emph{Proceedings of the 1st workshop on deep learning
  for recommender systems}}. \bibinfo{pages}{7--10}.
\newblock


\bibitem[\protect\citeauthoryear{Epure, Kille, Ingvaldsen, Deneckere, Salinesi,
  and Albayrak}{Epure et~al\mbox{.}}{2017}]%
        {epure_recommending_2017}
\bibfield{author}{\bibinfo{person}{Elena~Viorica Epure},
  \bibinfo{person}{Benjamin Kille}, \bibinfo{person}{Jon~Espen Ingvaldsen},
  \bibinfo{person}{Rebecca Deneckere}, \bibinfo{person}{Camille Salinesi},
  {and} \bibinfo{person}{Sahin Albayrak}.} \bibinfo{year}{2017}\natexlab{}.
\newblock \showarticletitle{Recommending personalized news in short user
  sessions}. In \bibinfo{booktitle}{\emph{Proceedings of the Eleventh ACM
  Conference on Recommender Systems}}. \bibinfo{pages}{121--129}.
\newblock


\bibitem[\protect\citeauthoryear{Gabriel De~Souza, Jannach, and
  Da~Cunha}{Gabriel De~Souza et~al\mbox{.}}{2019}]%
        {gabriel2019contextual}
\bibfield{author}{\bibinfo{person}{P~Moreira Gabriel De~Souza},
  \bibinfo{person}{Dietmar Jannach}, {and} \bibinfo{person}{Adilson~Marques
  Da~Cunha}.} \bibinfo{year}{2019}\natexlab{}.
\newblock \showarticletitle{Contextual hybrid session-based news recommendation
  with recurrent neural networks}.
\newblock \bibinfo{journal}{\emph{IEEE Access}}  \bibinfo{volume}{7}
  (\bibinfo{year}{2019}), \bibinfo{pages}{169185--169203}.
\newblock


\bibitem[\protect\citeauthoryear{Garg, Gupta, Malhotra, Vig, and Shroff}{Garg
  et~al\mbox{.}}{2019}]%
        {garg2019sequence}
\bibfield{author}{\bibinfo{person}{Diksha Garg}, \bibinfo{person}{Priyanka
  Gupta}, \bibinfo{person}{Pankaj Malhotra}, \bibinfo{person}{Lovekesh Vig},
  {and} \bibinfo{person}{Gautam Shroff}.} \bibinfo{year}{2019}\natexlab{}.
\newblock \showarticletitle{Sequence and time aware neighborhood for
  session-based recommendations: Stan}. In
  \bibinfo{booktitle}{\emph{Proceedings of the 42nd International ACM SIGIR
  Conference on Research and Development in Information Retrieval}}.
  \bibinfo{pages}{1069--1072}.
\newblock


\bibitem[\protect\citeauthoryear{Ge, Wu, Wu, Qi, and Huang}{Ge
  et~al\mbox{.}}{2020}]%
        {ge2020graph}
\bibfield{author}{\bibinfo{person}{Suyu Ge}, \bibinfo{person}{Chuhan Wu},
  \bibinfo{person}{Fangzhao Wu}, \bibinfo{person}{Tao Qi}, {and}
  \bibinfo{person}{Yongfeng Huang}.} \bibinfo{year}{2020}\natexlab{}.
\newblock \showarticletitle{Graph enhanced representation learning for news
  recommendation}. In \bibinfo{booktitle}{\emph{Proceedings of The Web
  Conference 2020}}. \bibinfo{pages}{2863--2869}.
\newblock


\bibitem[\protect\citeauthoryear{Gulla, Zhang, Liu, {\"O}zg{\"o}bek, and
  Su}{Gulla et~al\mbox{.}}{2017}]%
        {gulla_adressa_2017}
\bibfield{author}{\bibinfo{person}{Jon~Atle Gulla}, \bibinfo{person}{Lemei
  Zhang}, \bibinfo{person}{Peng Liu}, \bibinfo{person}{{\"O}zlem
  {\"O}zg{\"o}bek}, {and} \bibinfo{person}{Xiaomeng Su}.}
  \bibinfo{year}{2017}\natexlab{}.
\newblock \showarticletitle{The adressa dataset for news recommendation}. In
  \bibinfo{booktitle}{\emph{Proceedings of the international conference on web
  intelligence}}. \bibinfo{pages}{1042--1048}.
\newblock


\bibitem[\protect\citeauthoryear{Guo, Tang, Ye, Li, and He}{Guo
  et~al\mbox{.}}{2017}]%
        {guodeepfm2017}
\bibfield{author}{\bibinfo{person}{Huifeng Guo}, \bibinfo{person}{Ruiming
  Tang}, \bibinfo{person}{Yunming Ye}, \bibinfo{person}{Zhenguo Li}, {and}
  \bibinfo{person}{Xiuqiang He}.} \bibinfo{year}{2017}\natexlab{}.
\newblock \showarticletitle{DeepFM: a factorization-machine based neural
  network for CTR prediction}.
\newblock \bibinfo{journal}{\emph{arXiv preprint arXiv:1703.04247}}
  (\bibinfo{year}{2017}).
\newblock


\bibitem[\protect\citeauthoryear{Guo, Yin, Wang, Chen, Zhou, and Quoc
  Viet~Hung}{Guo et~al\mbox{.}}{2019}]%
        {guo_streaming_2019}
\bibfield{author}{\bibinfo{person}{Lei Guo}, \bibinfo{person}{Hongzhi Yin},
  \bibinfo{person}{Qinyong Wang}, \bibinfo{person}{Tong Chen},
  \bibinfo{person}{Alexander Zhou}, {and} \bibinfo{person}{Nguyen Quoc
  Viet~Hung}.} \bibinfo{year}{2019}\natexlab{}.
\newblock \showarticletitle{Streaming session-based recommendation}. In
  \bibinfo{booktitle}{\emph{Proceedings of the 25th ACM SIGKDD International
  Conference on Knowledge Discovery \& Data Mining}}.
  \bibinfo{pages}{1569--1577}.
\newblock


\bibitem[\protect\citeauthoryear{Hidasi and Karatzoglou}{Hidasi and
  Karatzoglou}{2018}]%
        {hidasi2018recurrent}
\bibfield{author}{\bibinfo{person}{Bal{\'a}zs Hidasi} {and}
  \bibinfo{person}{Alexandros Karatzoglou}.} \bibinfo{year}{2018}\natexlab{}.
\newblock \showarticletitle{Recurrent neural networks with top-k gains for
  session-based recommendations}. In \bibinfo{booktitle}{\emph{Proceedings of
  the 27th ACM International Conference on Information and Knowledge
  Management}}. \bibinfo{pages}{843--852}.
\newblock


\bibitem[\protect\citeauthoryear{Hidasi, Karatzoglou, Baltrunas, and
  Tikk}{Hidasi et~al\mbox{.}}{2015}]%
        {hidasi2015session}
\bibfield{author}{\bibinfo{person}{Bal{\'a}zs Hidasi},
  \bibinfo{person}{Alexandros Karatzoglou}, \bibinfo{person}{Linas Baltrunas},
  {and} \bibinfo{person}{Domonkos Tikk}.} \bibinfo{year}{2015}\natexlab{}.
\newblock \showarticletitle{Session-based recommendations with recurrent neural
  networks}.
\newblock \bibinfo{journal}{\emph{arXiv preprint arXiv:1511.06939}}
  (\bibinfo{year}{2015}).
\newblock


\bibitem[\protect\citeauthoryear{Hu, Xu, Li, Yang, Shi, Duan, Xie, and Zhou}{Hu
  et~al\mbox{.}}{2020}]%
        {hu2020graph}
\bibfield{author}{\bibinfo{person}{Linmei Hu}, \bibinfo{person}{Siyong Xu},
  \bibinfo{person}{Chen Li}, \bibinfo{person}{Cheng Yang},
  \bibinfo{person}{Chuan Shi}, \bibinfo{person}{Nan Duan},
  \bibinfo{person}{Xing Xie}, {and} \bibinfo{person}{Ming Zhou}.}
  \bibinfo{year}{2020}\natexlab{}.
\newblock \showarticletitle{Graph neural news recommendation with unsupervised
  preference disentanglement}. In \bibinfo{booktitle}{\emph{Proceedings of the
  58th Annual Meeting of the Association for Computational Linguistics}}.
  \bibinfo{pages}{4255--4264}.
\newblock


\bibitem[\protect\citeauthoryear{Jugovac, Jannach, and Karimi}{Jugovac
  et~al\mbox{.}}{2018}]%
        {jugovac_streamingrec:_2018}
\bibfield{author}{\bibinfo{person}{Michael Jugovac}, \bibinfo{person}{Dietmar
  Jannach}, {and} \bibinfo{person}{Mozhgan Karimi}.}
  \bibinfo{year}{2018}\natexlab{}.
\newblock \showarticletitle{Streamingrec: a framework for benchmarking
  stream-based news recommenders}. In \bibinfo{booktitle}{\emph{Proceedings of
  the 12th ACM Conference on Recommender Systems}}. \bibinfo{pages}{269--273}.
\newblock


\bibitem[\protect\citeauthoryear{Kaminskas and Bridge}{Kaminskas and
  Bridge}{2014}]%
        {kaminskas2014measuring}
\bibfield{author}{\bibinfo{person}{Marius Kaminskas} {and}
  \bibinfo{person}{Derek Bridge}.} \bibinfo{year}{2014}\natexlab{}.
\newblock \showarticletitle{Measuring surprise in recommender systems}. In
  \bibinfo{booktitle}{\emph{Proceedings of the workshop on recommender systems
  evaluation: dimensions and design (Workshop programme of the 8th ACM
  conference on recommender systems)}}. Citeseer.
\newblock


\bibitem[\protect\citeauthoryear{Kang and McAuley}{Kang and McAuley}{2018}]%
        {kang_self-attentive_2018}
\bibfield{author}{\bibinfo{person}{Wang-Cheng Kang} {and}
  \bibinfo{person}{Julian McAuley}.} \bibinfo{year}{2018}\natexlab{}.
\newblock \showarticletitle{Self-attentive sequential recommendation}. In
  \bibinfo{booktitle}{\emph{2018 IEEE International Conference on Data Mining
  (ICDM)}}. IEEE, \bibinfo{pages}{197--206}.
\newblock


\bibitem[\protect\citeauthoryear{Lee, Im, Jang, Cho, and Chung}{Lee
  et~al\mbox{.}}{2019}]%
        {lee_melu:_2019}
\bibfield{author}{\bibinfo{person}{Hoyeop Lee}, \bibinfo{person}{Jinbae Im},
  \bibinfo{person}{Seongwon Jang}, \bibinfo{person}{Hyunsouk Cho}, {and}
  \bibinfo{person}{Sehee Chung}.} \bibinfo{year}{2019}\natexlab{}.
\newblock \showarticletitle{Melu: Meta-learned user preference estimator for
  cold-start recommendation}. In \bibinfo{booktitle}{\emph{Proceedings of the
  25th ACM SIGKDD International Conference on Knowledge Discovery \& Data
  Mining}}. \bibinfo{pages}{1073--1082}.
\newblock


\bibitem[\protect\citeauthoryear{Lei, Ji, and Li}{Lei et~al\mbox{.}}{2019}]%
        {lei_tissa_2019}
\bibfield{author}{\bibinfo{person}{Chenyi Lei}, \bibinfo{person}{Shouling Ji},
  {and} \bibinfo{person}{Zhao Li}.} \bibinfo{year}{2019}\natexlab{}.
\newblock \showarticletitle{Tissa: A time slice self-attention approach for
  modeling sequential user behaviors}. In \bibinfo{booktitle}{\emph{The World
  Wide Web Conference}}. \bibinfo{pages}{2964--2970}.
\newblock


\bibitem[\protect\citeauthoryear{Li, Ren, Chen, Ren, Lian, and Ma}{Li
  et~al\mbox{.}}{2017}]%
        {li2017neural}
\bibfield{author}{\bibinfo{person}{Jing Li}, \bibinfo{person}{Pengjie Ren},
  \bibinfo{person}{Zhumin Chen}, \bibinfo{person}{Zhaochun Ren},
  \bibinfo{person}{Tao Lian}, {and} \bibinfo{person}{Jun Ma}.}
  \bibinfo{year}{2017}\natexlab{}.
\newblock \showarticletitle{Neural attentive session-based recommendation}. In
  \bibinfo{booktitle}{\emph{Proceedings of the 2017 ACM on Conference on
  Information and Knowledge Management}}. \bibinfo{pages}{1419--1428}.
\newblock


\bibitem[\protect\citeauthoryear{Lin, Cai, Wang, Liu, Zheng, and Shi}{Lin
  et~al\mbox{.}}{2020}]%
        {lin2020world}
\bibfield{author}{\bibinfo{person}{Zibo Lin}, \bibinfo{person}{Deng Cai},
  \bibinfo{person}{Yan Wang}, \bibinfo{person}{Xiaojiang Liu},
  \bibinfo{person}{Haitao Zheng}, {and} \bibinfo{person}{Shuming Shi}.}
  \bibinfo{year}{2020}\natexlab{}.
\newblock \showarticletitle{The World Is Not Binary: Learning to Rank with
  Grayscale Data for Dialogue Response Selection}. In
  \bibinfo{booktitle}{\emph{Proceedings of the 2020 Conference on Empirical
  Methods in Natural Language Processing (EMNLP)}}.
  \bibinfo{pages}{9220--9229}.
\newblock


\bibitem[\protect\citeauthoryear{Liu, Zeng, Mokhosi, and Zhang}{Liu
  et~al\mbox{.}}{2018}]%
        {liu2018stamp}
\bibfield{author}{\bibinfo{person}{Qiao Liu}, \bibinfo{person}{Yifu Zeng},
  \bibinfo{person}{Refuoe Mokhosi}, {and} \bibinfo{person}{Haibin Zhang}.}
  \bibinfo{year}{2018}\natexlab{}.
\newblock \showarticletitle{STAMP: short-term attention/memory priority model
  for session-based recommendation}. In \bibinfo{booktitle}{\emph{Proceedings
  of the 24th ACM SIGKDD International Conference on Knowledge Discovery \&
  Data Mining}}. \bibinfo{pages}{1831--1839}.
\newblock


\bibitem[\protect\citeauthoryear{Lu, Zhang, Ma, Shao, Liu, and Ma}{Lu
  et~al\mbox{.}}{2019}]%
        {lu_quality_2019}
\bibfield{author}{\bibinfo{person}{Hongyu Lu}, \bibinfo{person}{Min Zhang},
  \bibinfo{person}{Weizhi Ma}, \bibinfo{person}{Yunqiu Shao},
  \bibinfo{person}{Yiqun Liu}, {and} \bibinfo{person}{Shaoping Ma}.}
  \bibinfo{year}{2019}\natexlab{}.
\newblock \showarticletitle{Quality effects on user preferences and behaviorsin
  mobile news streaming}. In \bibinfo{booktitle}{\emph{The World Wide Web
  Conference}}. \bibinfo{pages}{1187--1197}.
\newblock


\bibitem[\protect\citeauthoryear{Ma, Kang, and Liu}{Ma et~al\mbox{.}}{2019}]%
        {ma2019hierarchical}
\bibfield{author}{\bibinfo{person}{Chen Ma}, \bibinfo{person}{Peng Kang}, {and}
  \bibinfo{person}{Xue Liu}.} \bibinfo{year}{2019}\natexlab{}.
\newblock \showarticletitle{Hierarchical gating networks for sequential
  recommendation}. In \bibinfo{booktitle}{\emph{Proceedings of the 25th ACM
  SIGKDD International Conference on Knowledge Discovery \& Data Mining}}.
  \bibinfo{pages}{825--833}.
\newblock


\bibitem[\protect\citeauthoryear{Moreira, Ferreira, and da~Cunha}{Moreira
  et~al\mbox{.}}{2018}]%
        {moreira_news_2018}
\bibfield{author}{\bibinfo{person}{Gabriel de Souza~P. Moreira},
  \bibinfo{person}{Felipe Ferreira}, {and} \bibinfo{person}{Adilson~Marques da
  Cunha}.} \bibinfo{year}{2018}\natexlab{}.
\newblock \showarticletitle{News {Session}-{Based} {Recommendations} using
  {Deep} {Neural} {Networks}}.
\newblock \bibinfo{journal}{\emph{Proceedings of the 3rd Workshop on Deep
  Learning for Recommender Systems - DLRS 2018}} (\bibinfo{year}{2018}),
  \bibinfo{pages}{15--23}.
\newblock
\newblock
\shownote{arXiv: 1808.00076.}


\bibitem[\protect\citeauthoryear{Pan, Cai, Chen, Chen, and de~Rijke}{Pan
  et~al\mbox{.}}{2020}]%
        {pan2020star}
\bibfield{author}{\bibinfo{person}{Zhiqiang Pan}, \bibinfo{person}{Fei Cai},
  \bibinfo{person}{Wanyu Chen}, \bibinfo{person}{Honghui Chen}, {and}
  \bibinfo{person}{Maarten de Rijke}.} \bibinfo{year}{2020}\natexlab{}.
\newblock \showarticletitle{Star Graph Neural Networks for Session-based
  Recommendation}. In \bibinfo{booktitle}{\emph{Proceedings of the 29th ACM
  International Conference on Information \& Knowledge Management}}.
  \bibinfo{pages}{1195--1204}.
\newblock


\bibitem[\protect\citeauthoryear{Pereira, Ueda, Penha, Santos, and
  Ziviani}{Pereira et~al\mbox{.}}{2019}]%
        {pereira2019online}
\bibfield{author}{\bibinfo{person}{Bruno~L Pereira}, \bibinfo{person}{Alberto
  Ueda}, \bibinfo{person}{Gustavo Penha}, \bibinfo{person}{Rodrygo~LT Santos},
  {and} \bibinfo{person}{Nivio Ziviani}.} \bibinfo{year}{2019}\natexlab{}.
\newblock \showarticletitle{Online learning to rank for sequential music
  recommendation}. In \bibinfo{booktitle}{\emph{Proceedings of the 13th ACM
  Conference on Recommender Systems}}. \bibinfo{pages}{237--245}.
\newblock


\bibitem[\protect\citeauthoryear{Rakkappan and Rajan}{Rakkappan and
  Rajan}{2019}]%
        {rakkappan2019context}
\bibfield{author}{\bibinfo{person}{Lakshmanan Rakkappan} {and}
  \bibinfo{person}{Vaibhav Rajan}.} \bibinfo{year}{2019}\natexlab{}.
\newblock \showarticletitle{Context-Aware Sequential Recommendations with
  Stacked Recurrent Neural Networks}. In \bibinfo{booktitle}{\emph{The World
  Wide Web Conference}}. \bibinfo{pages}{3172--3178}.
\newblock


\bibitem[\protect\citeauthoryear{Saunshi, Plevrakis, Arora, Khodak, and
  Khandeparkar}{Saunshi et~al\mbox{.}}{2019}]%
        {saunshi2019theoretical}
\bibfield{author}{\bibinfo{person}{Nikunj Saunshi}, \bibinfo{person}{Orestis
  Plevrakis}, \bibinfo{person}{Sanjeev Arora}, \bibinfo{person}{Mikhail
  Khodak}, {and} \bibinfo{person}{Hrishikesh Khandeparkar}.}
  \bibinfo{year}{2019}\natexlab{}.
\newblock \showarticletitle{A theoretical analysis of contrastive unsupervised
  representation learning}. In \bibinfo{booktitle}{\emph{International
  Conference on Machine Learning}}. PMLR, \bibinfo{pages}{5628--5637}.
\newblock


\bibitem[\protect\citeauthoryear{Sheu, Chu, Qi, and Li}{Sheu
  et~al\mbox{.}}{2021}]%
        {sheu2021knowledge}
\bibfield{author}{\bibinfo{person}{Heng-Shiou Sheu}, \bibinfo{person}{Zhixuan
  Chu}, \bibinfo{person}{Daiqing Qi}, {and} \bibinfo{person}{Sheng Li}.}
  \bibinfo{year}{2021}\natexlab{}.
\newblock \showarticletitle{Knowledge-guided article embedding refinement for
  session-based news recommendation}.
\newblock \bibinfo{journal}{\emph{IEEE Transactions on Neural Networks and
  Learning Systems}} (\bibinfo{year}{2021}).
\newblock


\bibitem[\protect\citeauthoryear{Sheu and Li}{Sheu and Li}{2020}]%
        {sheu2020context}
\bibfield{author}{\bibinfo{person}{Heng-Shiou Sheu} {and}
  \bibinfo{person}{Sheng Li}.} \bibinfo{year}{2020}\natexlab{}.
\newblock \showarticletitle{Context-aware graph embedding for session-based
  news recommendation}. In \bibinfo{booktitle}{\emph{Fourteenth ACM conference
  on recommender systems}}. \bibinfo{pages}{657--662}.
\newblock


\bibitem[\protect\citeauthoryear{Smadja, Grusky, Artzi, and Naaman}{Smadja
  et~al\mbox{.}}{2019}]%
        {smadja_understanding_2019}
\bibfield{author}{\bibinfo{person}{Uzi Smadja}, \bibinfo{person}{Max Grusky},
  \bibinfo{person}{Yoav Artzi}, {and} \bibinfo{person}{Mor Naaman}.}
  \bibinfo{year}{2019}\natexlab{}.
\newblock \showarticletitle{Understanding reader backtracking behavior in
  online news articles}. In \bibinfo{booktitle}{\emph{The World Wide Web
  Conference}}. \bibinfo{pages}{3237--3243}.
\newblock


\bibitem[\protect\citeauthoryear{Song, Shen, Ou, Zhang, Xiao, and Liang}{Song
  et~al\mbox{.}}{2019}]%
        {song_islf_2019}
\bibfield{author}{\bibinfo{person}{Jing Song}, \bibinfo{person}{Hong Shen},
  \bibinfo{person}{Zijing Ou}, \bibinfo{person}{Junyi Zhang},
  \bibinfo{person}{Teng Xiao}, {and} \bibinfo{person}{Shangsong Liang}.}
  \bibinfo{year}{2019}\natexlab{}.
\newblock \showarticletitle{ISLF: Interest Shift and Latent Factors Combination
  Model for Session-based Recommendation.}. In
  \bibinfo{booktitle}{\emph{IJCAI}}. \bibinfo{pages}{5765--5771}.
\newblock


\bibitem[\protect\citeauthoryear{Sottocornola, Symeonidis, and
  Zanker}{Sottocornola et~al\mbox{.}}{2018}]%
        {sottocornola2018session}
\bibfield{author}{\bibinfo{person}{Gabriele Sottocornola},
  \bibinfo{person}{Panagiotis Symeonidis}, {and} \bibinfo{person}{Markus
  Zanker}.} \bibinfo{year}{2018}\natexlab{}.
\newblock \showarticletitle{Session-based news recommendations}. In
  \bibinfo{booktitle}{\emph{Companion Proceedings of the The Web Conference
  2018}}. \bibinfo{pages}{1395--1399}.
\newblock


\bibitem[\protect\citeauthoryear{Symeonidis, Kirjackaja, and Zanker}{Symeonidis
  et~al\mbox{.}}{2020}]%
        {symeonidis2020session}
\bibfield{author}{\bibinfo{person}{Panagiotis Symeonidis},
  \bibinfo{person}{Lidija Kirjackaja}, {and} \bibinfo{person}{Markus Zanker}.}
  \bibinfo{year}{2020}\natexlab{}.
\newblock \showarticletitle{Session-aware news recommendations using random
  walks on time-evolving heterogeneous information networks}.
\newblock \bibinfo{journal}{\emph{User Modeling and User-Adapted Interaction}}
  (\bibinfo{year}{2020}), \bibinfo{pages}{1--29}.
\newblock


\bibitem[\protect\citeauthoryear{Symeonidis, Kirjackaja, and Zanker}{Symeonidis
  et~al\mbox{.}}{2021}]%
        {symeonidis2021session}
\bibfield{author}{\bibinfo{person}{Panagiotis Symeonidis},
  \bibinfo{person}{Lidija Kirjackaja}, {and} \bibinfo{person}{Markus Zanker}.}
  \bibinfo{year}{2021}\natexlab{}.
\newblock \showarticletitle{Session-based news recommendations using SimRank on
  multi-modal graphs}.
\newblock \bibinfo{journal}{\emph{Expert Systems with Applications}}
  \bibinfo{volume}{180} (\bibinfo{year}{2021}), \bibinfo{pages}{115028}.
\newblock


\bibitem[\protect\citeauthoryear{Vaswani, Shazeer, Parmar, Uszkoreit, Jones,
  Gomez, Kaiser, and Polosukhin}{Vaswani et~al\mbox{.}}{2017}]%
        {vaswani2017attention}
\bibfield{author}{\bibinfo{person}{Ashish Vaswani}, \bibinfo{person}{Noam
  Shazeer}, \bibinfo{person}{Niki Parmar}, \bibinfo{person}{Jakob Uszkoreit},
  \bibinfo{person}{Llion Jones}, \bibinfo{person}{Aidan~N Gomez},
  \bibinfo{person}{{\L}ukasz Kaiser}, {and} \bibinfo{person}{Illia
  Polosukhin}.} \bibinfo{year}{2017}\natexlab{}.
\newblock \showarticletitle{Attention is all you need}. In
  \bibinfo{booktitle}{\emph{Advances in neural information processing
  systems}}. \bibinfo{pages}{5998--6008}.
\newblock


\bibitem[\protect\citeauthoryear{Wang, Zhang, Ma, Liu, and Ma}{Wang
  et~al\mbox{.}}{2020c}]%
        {wang2020make}
\bibfield{author}{\bibinfo{person}{Chenyang Wang}, \bibinfo{person}{Min Zhang},
  \bibinfo{person}{Weizhi Ma}, \bibinfo{person}{Yiqun Liu}, {and}
  \bibinfo{person}{Shaoping Ma}.} \bibinfo{year}{2020}\natexlab{c}.
\newblock \showarticletitle{Make it a chorus: knowledge-and time-aware item
  modeling for sequential recommendation}. In
  \bibinfo{booktitle}{\emph{Proceedings of the 43rd International ACM SIGIR
  Conference on Research and Development in Information Retrieval}}.
  \bibinfo{pages}{109--118}.
\newblock


\bibitem[\protect\citeauthoryear{Wang, Wu, Liu, and Xie}{Wang
  et~al\mbox{.}}{2020b}]%
        {wang2020fine}
\bibfield{author}{\bibinfo{person}{Heyuan Wang}, \bibinfo{person}{Fangzhao Wu},
  \bibinfo{person}{Zheng Liu}, {and} \bibinfo{person}{Xing Xie}.}
  \bibinfo{year}{2020}\natexlab{b}.
\newblock \showarticletitle{Fine-grained Interest Matching for Neural News
  Recommendation}. In \bibinfo{booktitle}{\emph{Proceedings of the 58th Annual
  Meeting of the Association for Computational Linguistics}}.
  \bibinfo{pages}{836--845}.
\newblock


\bibitem[\protect\citeauthoryear{Wang, Zhang, Wang, Zhao, Li, Xie, and
  Guo}{Wang et~al\mbox{.}}{2018c}]%
        {wang_ripplenet:_2018}
\bibfield{author}{\bibinfo{person}{Hongwei Wang}, \bibinfo{person}{Fuzheng
  Zhang}, \bibinfo{person}{Jialin Wang}, \bibinfo{person}{Miao Zhao},
  \bibinfo{person}{Wenjie Li}, \bibinfo{person}{Xing Xie}, {and}
  \bibinfo{person}{Minyi Guo}.} \bibinfo{year}{2018}\natexlab{c}.
\newblock \showarticletitle{Ripplenet: Propagating user preferences on the
  knowledge graph for recommender systems}. In
  \bibinfo{booktitle}{\emph{Proceedings of the 27th ACM International
  Conference on Information and Knowledge Management}}.
  \bibinfo{pages}{417--426}.
\newblock


\bibitem[\protect\citeauthoryear{Wang, Zhang, Xie, and Guo}{Wang
  et~al\mbox{.}}{2018d}]%
        {wang2018dkn}
\bibfield{author}{\bibinfo{person}{Hongwei Wang}, \bibinfo{person}{Fuzheng
  Zhang}, \bibinfo{person}{Xing Xie}, {and} \bibinfo{person}{Minyi Guo}.}
  \bibinfo{year}{2018}\natexlab{d}.
\newblock \showarticletitle{DKN: Deep knowledge-aware network for news
  recommendation}. In \bibinfo{booktitle}{\emph{Proceedings of the 2018 world
  wide web conference}}. \bibinfo{pages}{1835--1844}.
\newblock


\bibitem[\protect\citeauthoryear{Wang, Gong, Zheng, and Zhang}{Wang
  et~al\mbox{.}}{2018a}]%
        {wang2018modeling}
\bibfield{author}{\bibinfo{person}{Menghan Wang}, \bibinfo{person}{Mingming
  Gong}, \bibinfo{person}{Xiaolin Zheng}, {and} \bibinfo{person}{Kun Zhang}.}
  \bibinfo{year}{2018}\natexlab{a}.
\newblock \showarticletitle{Modeling dynamic missingness of implicit feedback
  for recommendation}.
\newblock \bibinfo{journal}{\emph{Advances in neural information processing
  systems}}  \bibinfo{volume}{31} (\bibinfo{year}{2018}),
  \bibinfo{pages}{6669}.
\newblock


\bibitem[\protect\citeauthoryear{Wang, Yin, Hu, Lian, Wang, and Huang}{Wang
  et~al\mbox{.}}{2018b}]%
        {wang_neural_2018}
\bibfield{author}{\bibinfo{person}{Qinyong Wang}, \bibinfo{person}{Hongzhi
  Yin}, \bibinfo{person}{Zhiting Hu}, \bibinfo{person}{Defu Lian},
  \bibinfo{person}{Hao Wang}, {and} \bibinfo{person}{Zi Huang}.}
  \bibinfo{year}{2018}\natexlab{b}.
\newblock \showarticletitle{Neural memory streaming recommender networks with
  adversarial training}. In \bibinfo{booktitle}{\emph{Proceedings of the 24th
  ACM SIGKDD International Conference on Knowledge Discovery \& Data Mining}}.
  \bibinfo{pages}{2467--2475}.
\newblock


\bibitem[\protect\citeauthoryear{Wang, Hu, Wang, Sheng, Orgun, and Cao}{Wang
  et~al\mbox{.}}{2019}]%
        {wang2019modeling}
\bibfield{author}{\bibinfo{person}{Shoujin Wang}, \bibinfo{person}{Liang Hu},
  \bibinfo{person}{Yang Wang}, \bibinfo{person}{Quan~Z Sheng},
  \bibinfo{person}{Mehmet Orgun}, {and} \bibinfo{person}{Longbing Cao}.}
  \bibinfo{year}{2019}\natexlab{}.
\newblock \showarticletitle{Modeling multi-purpose sessions for nextitem
  recommendations via mixture-channel purpose routing networks}. In
  \bibinfo{booktitle}{\emph{Proceedings of the 28th International Joint
  Conference on Artificial Intelligence}}. AAAI Press, \bibinfo{pages}{1--7}.
\newblock


\bibitem[\protect\citeauthoryear{Wang, Feng, He, Zhang, and Chua}{Wang
  et~al\mbox{.}}{2020a}]%
        {wang2020click}
\bibfield{author}{\bibinfo{person}{Wenjie Wang}, \bibinfo{person}{Fuli Feng},
  \bibinfo{person}{Xiangnan He}, \bibinfo{person}{Hanwang Zhang}, {and}
  \bibinfo{person}{Tat-Seng Chua}.} \bibinfo{year}{2020}\natexlab{a}.
\newblock \showarticletitle{``Click'' Is Not Equal to ``Like'': Counterfactual
  Recommendation for Mitigating Clickbait Issue}.
\newblock \bibinfo{journal}{\emph{arXiv preprint arXiv:2009.09945}}
  (\bibinfo{year}{2020}).
\newblock


\bibitem[\protect\citeauthoryear{Wu, Wu, An, Huang, Huang, and Xie}{Wu
  et~al\mbox{.}}{2019b}]%
        {wu2019npa}
\bibfield{author}{\bibinfo{person}{Chuhan Wu}, \bibinfo{person}{Fangzhao Wu},
  \bibinfo{person}{Mingxiao An}, \bibinfo{person}{Jianqiang Huang},
  \bibinfo{person}{Yongfeng Huang}, {and} \bibinfo{person}{Xing Xie}.}
  \bibinfo{year}{2019}\natexlab{b}.
\newblock \showarticletitle{Npa: Neural news recommendation with personalized
  attention}. In \bibinfo{booktitle}{\emph{Proceedings of the 25th ACM SIGKDD
  International Conference on Knowledge Discovery \& Data Mining}}.
  \bibinfo{pages}{2576--2584}.
\newblock


\bibitem[\protect\citeauthoryear{Wu, Wu, An, Qi, Huang, Huang, and Xie}{Wu
  et~al\mbox{.}}{2019c}]%
        {wu_neural_2019-1}
\bibfield{author}{\bibinfo{person}{Chuhan Wu}, \bibinfo{person}{Fangzhao Wu},
  \bibinfo{person}{Mingxiao An}, \bibinfo{person}{Tao Qi},
  \bibinfo{person}{Jianqiang Huang}, \bibinfo{person}{Yongfeng Huang}, {and}
  \bibinfo{person}{Xing Xie}.} \bibinfo{year}{2019}\natexlab{c}.
\newblock \showarticletitle{Neural news recommendation with heterogeneous user
  behavior}. In \bibinfo{booktitle}{\emph{Proceedings of the 2019 Conference on
  Empirical Methods in Natural Language Processing and the 9th International
  Joint Conference on Natural Language Processing (EMNLP-IJCNLP)}}.
  \bibinfo{pages}{4874--4883}.
\newblock


\bibitem[\protect\citeauthoryear{Wu, Wu, Qi, and Huang}{Wu
  et~al\mbox{.}}{2020c}]%
        {wu2020CPRS}
\bibfield{author}{\bibinfo{person}{Chuhan Wu}, \bibinfo{person}{Fangzhao Wu},
  \bibinfo{person}{Tao Qi}, {and} \bibinfo{person}{Yongfeng Huang}.}
  \bibinfo{year}{2020}\natexlab{c}.
\newblock \showarticletitle{User Modeling with Click Preference and Reading
  Satisfaction for News Recommendation}. In
  \bibinfo{booktitle}{\emph{Proceedings of the Twenty-Ninth International Joint
  Conference on Artificial Intelligence, {IJCAI-20}}}.
  \bibinfo{pages}{3023--3029}.
\newblock


\bibitem[\protect\citeauthoryear{Wu, Qiao, Chen, Wu, Qi, Lian, Liu, Xie, Gao,
  Wu, et~al\mbox{.}}{Wu et~al\mbox{.}}{2020b}]%
        {wu2020mind}
\bibfield{author}{\bibinfo{person}{Fangzhao Wu}, \bibinfo{person}{Ying Qiao},
  \bibinfo{person}{Jiun-Hung Chen}, \bibinfo{person}{Chuhan Wu},
  \bibinfo{person}{Tao Qi}, \bibinfo{person}{Jianxun Lian},
  \bibinfo{person}{Danyang Liu}, \bibinfo{person}{Xing Xie},
  \bibinfo{person}{Jianfeng Gao}, \bibinfo{person}{Winnie Wu}, {et~al\mbox{.}}}
  \bibinfo{year}{2020}\natexlab{b}.
\newblock \showarticletitle{Mind: A large-scale dataset for news
  recommendation}. In \bibinfo{booktitle}{\emph{Proceedings of the 58th Annual
  Meeting of the Association for Computational Linguistics}}.
  \bibinfo{pages}{3597--3606}.
\newblock


\bibitem[\protect\citeauthoryear{Wu, Cai, and Wang}{Wu et~al\mbox{.}}{2020a}]%
        {wu2020deja}
\bibfield{author}{\bibinfo{person}{Jibang Wu}, \bibinfo{person}{Renqin Cai},
  {and} \bibinfo{person}{Hongning Wang}.} \bibinfo{year}{2020}\natexlab{a}.
\newblock \showarticletitle{D{\'e}j{\`a} vu: A Contextualized Temporal
  Attention Mechanism for Sequential Recommendation}. In
  \bibinfo{booktitle}{\emph{Proceedings of The Web Conference 2020}}.
  \bibinfo{pages}{2199--2209}.
\newblock


\bibitem[\protect\citeauthoryear{Wu, Zhu, Yu, Rajendra, Zhao, Aghdaie, and
  Zaman}{Wu et~al\mbox{.}}{2019d}]%
        {wu_recommender_2019}
\bibfield{author}{\bibinfo{person}{Meng Wu}, \bibinfo{person}{Ying Zhu},
  \bibinfo{person}{Qilian Yu}, \bibinfo{person}{Bhargav Rajendra},
  \bibinfo{person}{Yunqi Zhao}, \bibinfo{person}{Navid Aghdaie}, {and}
  \bibinfo{person}{Kazi~A Zaman}.} \bibinfo{year}{2019}\natexlab{d}.
\newblock \showarticletitle{A recommender system for heterogeneous and time
  sensitive environment}. In \bibinfo{booktitle}{\emph{Proceedings of the 13th
  ACM Conference on Recommender Systems}}. \bibinfo{pages}{210--218}.
\newblock


\bibitem[\protect\citeauthoryear{Wu, Tang, Zhu, Wang, Xie, and Tan}{Wu
  et~al\mbox{.}}{2019a}]%
        {wu2019session}
\bibfield{author}{\bibinfo{person}{Shu Wu}, \bibinfo{person}{Yuyuan Tang},
  \bibinfo{person}{Yanqiao Zhu}, \bibinfo{person}{Liang Wang},
  \bibinfo{person}{Xing Xie}, {and} \bibinfo{person}{Tieniu Tan}.}
  \bibinfo{year}{2019}\natexlab{a}.
\newblock \showarticletitle{Session-based recommendation with graph neural
  networks}. In \bibinfo{booktitle}{\emph{Proceedings of the AAAI Conference on
  Artificial Intelligence}}, Vol.~\bibinfo{volume}{33}.
  \bibinfo{pages}{346--353}.
\newblock


\bibitem[\protect\citeauthoryear{Xiao, Liang, and Meng}{Xiao
  et~al\mbox{.}}{2019}]%
        {xiao2019hierarchical}
\bibfield{author}{\bibinfo{person}{Teng Xiao}, \bibinfo{person}{Shangsong
  Liang}, {and} \bibinfo{person}{Zaiqiao Meng}.}
  \bibinfo{year}{2019}\natexlab{}.
\newblock \showarticletitle{Hierarchical neural variational model for
  personalized sequential recommendation}. In \bibinfo{booktitle}{\emph{The
  World Wide Web Conference}}. \bibinfo{pages}{3377--3383}.
\newblock


\bibitem[\protect\citeauthoryear{Xie, Ling, Wang, Wang, Xia, and Lin}{Xie
  et~al\mbox{.}}{2020}]%
        {xie2020deep}
\bibfield{author}{\bibinfo{person}{Ruobing Xie}, \bibinfo{person}{Cheng Ling},
  \bibinfo{person}{Yalong Wang}, \bibinfo{person}{Rui Wang},
  \bibinfo{person}{Feng Xia}, {and} \bibinfo{person}{Leyu Lin}.}
  \bibinfo{year}{2020}\natexlab{}.
\newblock \showarticletitle{Deep Feedback Network for Recommendation}.
\newblock \bibinfo{journal}{\emph{Proceedings of IJCAI-PRICAI}}
  (\bibinfo{year}{2020}).
\newblock


\bibitem[\protect\citeauthoryear{Xu, Zhao, Liu, Sheng, Xu, Zhuang, Fang, and
  Zhou}{Xu et~al\mbox{.}}{2019b}]%
        {xu2019graph}
\bibfield{author}{\bibinfo{person}{Chengfeng Xu}, \bibinfo{person}{Pengpeng
  Zhao}, \bibinfo{person}{Yanchi Liu}, \bibinfo{person}{Victor~S Sheng},
  \bibinfo{person}{Jiajie Xu}, \bibinfo{person}{Fuzhen Zhuang},
  \bibinfo{person}{Junhua Fang}, {and} \bibinfo{person}{Xiaofang Zhou}.}
  \bibinfo{year}{2019}\natexlab{b}.
\newblock \showarticletitle{Graph contextualized self-attention network for
  session-based recommendation}. In \bibinfo{booktitle}{\emph{Proc. 28th Int.
  Joint Conf. Artif. Intell.(IJCAI)}}. \bibinfo{pages}{3940--3946}.
\newblock


\bibitem[\protect\citeauthoryear{Xu, Zhao, Liu, Xu, S.~Sheng, Cui, Zhou, and
  Xiong}{Xu et~al\mbox{.}}{2019c}]%
        {xu_recurrent_2019}
\bibfield{author}{\bibinfo{person}{Chengfeng Xu}, \bibinfo{person}{Pengpeng
  Zhao}, \bibinfo{person}{Yanchi Liu}, \bibinfo{person}{Jiajie Xu},
  \bibinfo{person}{Victor S~Sheng S.~Sheng}, \bibinfo{person}{Zhiming Cui},
  \bibinfo{person}{Xiaofang Zhou}, {and} \bibinfo{person}{Hui Xiong}.}
  \bibinfo{year}{2019}\natexlab{c}.
\newblock \showarticletitle{Recurrent convolutional neural network for
  sequential recommendation}. In \bibinfo{booktitle}{\emph{The world wide web
  conference}}. \bibinfo{pages}{3398--3404}.
\newblock


\bibitem[\protect\citeauthoryear{Xu, Xu, Chen, Han, Li, Tan, Shen, and Shen}{Xu
  et~al\mbox{.}}{2019a}]%
        {xu2019time}
\bibfield{author}{\bibinfo{person}{Qidi Xu}, \bibinfo{person}{Haocheng Xu},
  \bibinfo{person}{Weilong Chen}, \bibinfo{person}{Chaojun Han},
  \bibinfo{person}{Haoyang Li}, \bibinfo{person}{Wenxin Tan},
  \bibinfo{person}{Fumin Shen}, {and} \bibinfo{person}{Heng~Tao Shen}.}
  \bibinfo{year}{2019}\natexlab{a}.
\newblock \showarticletitle{Time-aware Session Embedding for Click-Through-Rate
  Prediction}. In \bibinfo{booktitle}{\emph{Proceedings of the 27th ACM
  International Conference on Multimedia}}. \bibinfo{pages}{2617--2621}.
\newblock


\bibitem[\protect\citeauthoryear{Zhang, Liu, and Gulla}{Zhang
  et~al\mbox{.}}{2018}]%
        {zhang2018deep}
\bibfield{author}{\bibinfo{person}{Lemei Zhang}, \bibinfo{person}{Peng Liu},
  {and} \bibinfo{person}{Jon~Atle Gulla}.} \bibinfo{year}{2018}\natexlab{}.
\newblock \showarticletitle{A deep joint network for session-based news
  recommendations with contextual augmentation}.
\newblock In \bibinfo{booktitle}{\emph{Proceedings of the 29th on Hypertext and
  Social Media}}. \bibinfo{pages}{201--209}.
\newblock


\bibitem[\protect\citeauthoryear{Zhang, Liu, and Gulla}{Zhang
  et~al\mbox{.}}{2019a}]%
        {zhang2019dynamic}
\bibfield{author}{\bibinfo{person}{Lemei Zhang}, \bibinfo{person}{Peng Liu},
  {and} \bibinfo{person}{Jon~Atle Gulla}.} \bibinfo{year}{2019}\natexlab{a}.
\newblock \showarticletitle{Dynamic attention-integrated neural network for
  session-based news recommendation}.
\newblock \bibinfo{journal}{\emph{Machine Learning}} \bibinfo{volume}{108},
  \bibinfo{number}{10} (\bibinfo{year}{2019}), \bibinfo{pages}{1851--1875}.
\newblock


\bibitem[\protect\citeauthoryear{Zhang, Zhao, Liu, Sheng, Xu, Wang, Liu, and
  Zhou}{Zhang et~al\mbox{.}}{2019b}]%
        {zhang_feature-level_2019}
\bibfield{author}{\bibinfo{person}{Tingting Zhang}, \bibinfo{person}{Pengpeng
  Zhao}, \bibinfo{person}{Yanchi Liu}, \bibinfo{person}{Victor~S Sheng},
  \bibinfo{person}{Jiajie Xu}, \bibinfo{person}{Deqing Wang},
  \bibinfo{person}{Guanfeng Liu}, {and} \bibinfo{person}{Xiaofang Zhou}.}
  \bibinfo{year}{2019}\natexlab{b}.
\newblock \showarticletitle{Feature-level Deeper Self-Attention Network for
  Sequential Recommendation.}. In \bibinfo{booktitle}{\emph{IJCAI}}.
  \bibinfo{pages}{4320--4326}.
\newblock


\bibitem[\protect\citeauthoryear{Zheng, Zhang, Zheng, Xiang, Yuan, Xie, and
  Li}{Zheng et~al\mbox{.}}{2018}]%
        {zheng2018drn}
\bibfield{author}{\bibinfo{person}{Guanjie Zheng}, \bibinfo{person}{Fuzheng
  Zhang}, \bibinfo{person}{Zihan Zheng}, \bibinfo{person}{Yang Xiang},
  \bibinfo{person}{Nicholas~Jing Yuan}, \bibinfo{person}{Xing Xie}, {and}
  \bibinfo{person}{Zhenhui Li}.} \bibinfo{year}{2018}\natexlab{}.
\newblock \showarticletitle{DRN: A deep reinforcement learning framework for
  news recommendation}. In \bibinfo{booktitle}{\emph{Proceedings of the 2018
  World Wide Web Conference}}. \bibinfo{pages}{167--176}.
\newblock


\bibitem[\protect\citeauthoryear{Zhou, Wen, Zhang, Trajcevski, and Zhong}{Zhou
  et~al\mbox{.}}{2019}]%
        {zhou_variational_2019}
\bibfield{author}{\bibinfo{person}{Fan Zhou}, \bibinfo{person}{Zijing Wen},
  \bibinfo{person}{Kunpeng Zhang}, \bibinfo{person}{Goce Trajcevski}, {and}
  \bibinfo{person}{Ting Zhong}.} \bibinfo{year}{2019}\natexlab{}.
\newblock \showarticletitle{Variational session-based recommendation using
  normalizing flows}. In \bibinfo{booktitle}{\emph{The World Wide Web
  Conference}}. \bibinfo{pages}{3476--3475}.
\newblock


\bibitem[\protect\citeauthoryear{Zhu, Zhou, Song, Tan, and Guo}{Zhu
  et~al\mbox{.}}{2019}]%
        {zhu2019dan}
\bibfield{author}{\bibinfo{person}{Qiannan Zhu}, \bibinfo{person}{Xiaofei
  Zhou}, \bibinfo{person}{Zeliang Song}, \bibinfo{person}{Jianlong Tan}, {and}
  \bibinfo{person}{Li Guo}.} \bibinfo{year}{2019}\natexlab{}.
\newblock \showarticletitle{Dan: Deep attention neural network for news
  recommendation}. In \bibinfo{booktitle}{\emph{Proceedings of the AAAI
  Conference on Artificial Intelligence}}, Vol.~\bibinfo{volume}{33}.
  \bibinfo{pages}{5973--5980}.
\newblock


\end{thebibliography}




\end{document}